\RequirePackage{fixltx2e}
\documentclass[aps,prx,reprint,twocolumn,amsmath,amssymb,amsfonts,longbibliography,superscriptaddress]{revtex4-1}
\usepackage{tikz}
\usepackage{tikz-cd}
\usepackage{amsmath}
\usepackage{times}
\usepackage[varg]{txfonts}
\usepackage{hyperref}
\usepackage{color}
\usepackage{pdfpages}

\usepackage{mathrsfs}
\usepackage{color}
\usepackage{graphicx}
\usepackage[percent]{overpic}
\usepackage{bm}
\usepackage{fixltx2e}
\usepackage{natbib}

\renewcommand{\vec}[1]{\mathbf{\bm{#1}}}
\newcommand{\vhat}[1]{\vec{\hat{#1}}}
\newcommand{\mat}[1]{\mathbf{\bm{#1}}}

\newcommand{\bra}[1]{\left \langle #1\right|} 
\newcommand{\brap}[1]{\langle #1 |} 
\newcommand{\ketp}[1]{|#1 \rangle} \newcommand{\ket}[1]{\left| #1  \right \rangle}

\newcommand{\braketp}[2]{\langle  #1 | #2 \rangle}
\newcommand{\exc}[3]{\left< #1 \vphantom{#2#3} \right| #2 \left| #3 \vphantom{#1#2} \right>}

\newcommand{\avg}[1]{\langle #1 \rangle}

\newcommand{\h}[1]{{#1}^{\dagger}} 
\newcommand{\trp}[1]{{#1}^{\intercal}} 
\newcommand{\cc}[1]{{#1}^{*}}\newcommand{\cb}[1]{\bar{#1}}

\newcommand{\up}{\uparrow}
\newcommand{\down}{\downarrow}

\newcommand{\im}{{\rm Im}}

\newcommand{\tr}{{\rm tr}}
\newcommand{\meV}{\ {\rm meV}}
\newcommand{\mueV}{\ {\rm \mu eV}}
\newcommand{\K}{\ {\rm K}}

\newcommand{\tsup}[1]{\textsuperscript{#1}}
\newcommand{\tsub}[1]{\textsubscript{#1}}

\newcommand{\abo}[2]{#1\tsub{2}#2\tsub{2}O\tsub{7}}
\newcommand{\eto}{\abo{Er}{Ti}}
\newcommand{\hto}{\abo{Ho}{Ti}}
\newcommand{\yto}{\abo{Yb}{Ti}}
\newcommand{\tto}{\abo{Tb}{Ti}}
\newcommand{\rto}[1]{\abo{#1}{Ti}}

\newcommand{\mJ}{J}
\newcommand{\mS}{S}

\newcommand{\exJ}{\mathcal{J}}
\newcommand{\exM}{\mathcal{M}}
\newcommand{\exL}{\mathcal{L}}
\newcommand{\exK}{\mathcal{K}}
\newcommand{\cfV}{\mathcal{V}}

\newcommand{\Heff}[1]{H_{\rm eff,#1}}

\newcommand{\exJpm}{\exJ_{\pm}}
\newcommand{\exJpp}{\exJ_{\pm\pm}}
\newcommand{\exJzp}{\exJ_{z\pm}}
\newcommand{\exJzz}{\exJ_{zz}}

\newcommand{\cefS}{\Lambda}
\newcommand{\cefX}[1]{(\eta \cefS)^{#1}}

\definecolor{cred}{RGB}{228,26,28}
\definecolor{cblue}{RGB}{55,126,184}
\definecolor{clblue}{RGB}{205,223,237}
\definecolor{cgreen}{RGB}{77,175,74}
\definecolor{cgray}{RGB}{150,150,150}
\definecolor{clgray}{RGB}{200,200,200}
\definecolor{cpurple}{RGB}{152,78,163}
\definecolor{corange}{RGB}{255,127,0}
\definecolor{cgold}{RGB}{230,171,2}

\hypersetup{colorlinks=true,linkcolor=cblue,citecolor=cblue,urlcolor=cblue} 

\begin{document}
 
\title{Order by virtual crystal field fluctuations in pyrochlore XY antiferromagnets}
\author{Jeffrey G. Rau}
\affiliation{Department of Physics and Astronomy, University of
Waterloo, Ontario, N2L 3G1, Canada} 
\author{Sylvain Petit}
\affiliation{CEA, Centre de Saclay, DSM/IRAMIS/ Laboratoire L´eon
Brillouin, F-91191 Gif-sur-Yvette, France} 
\author{Michel J. P. Gingras} 
\affiliation{Department of Physics and Astronomy,
University of Waterloo, Ontario, N2L 3G1, Canada}
\affiliation{Perimeter Institute for Theoretical Physics, Waterloo,
Ontario, N2L 2Y5, Canada} 
\affiliation{Canadian Institute for Advanced
Research, 180 Dundas Street West, Suite 1400, Toronto, ON, M5G 1Z8,
Canada}
\affiliation{
Quantum Matter Institute, University of British Columbia, Vancouver, BC, V6T 1Z4, Canada
}
\affiliation{
TRIUMF, Theory Group, 4004 Wesbrook Mall, Vancouver, BC V6T 2A3, Canada
}
\date{\today}

\begin{abstract}
Conclusive evidence of order by disorder is scarce in real
materials. Perhaps one of the strongest cases presented has been for the
pyrochlore XY antiferromagnet \eto{}, with the ground state selection
proceeding by order by disorder induced through the effects of quantum
fluctuations. This identification assumes the smallness of the
effect of virtual crystal field fluctuations that could provide an
alternative route to picking the ground state.  Here we show that this
\emph{order by virtual crystal field fluctuations} is not only
significant, but competitive with the effects of quantum fluctuations.
Further, we argue that higher-multipolar interactions that are
generically present in rare-earth magnets can dramatically enhance
this effect. From a simplified bilinear-biquadratic model of these
multipolar interactions, we show how the virtual crystal field
fluctuations manifest in \eto{} using a combination of strong coupling
perturbation theory and the random phase approximation. We find that
the experimentally observed $\psi_2$ state is indeed selected and the experimentally
measured excitation gap can be reproduced when the bilinear and
biquadratic couplings are comparable while maintaining agreement with
the entire experimental spin-wave excitation spectrum.  Finally,
we comments on possible tests of this scenario and discuss
implications for other order-by-disorder candidates in rare-earth
magnets.
\end{abstract}

\maketitle

\section{Introduction}
The study of frustrated magnetism has led to the discovery of several
new and exotic phenomena \cite{springer-2011-frustrated}. Much of this physics
can be traced back to frustration inducing a large number
of degenerate or nearly degenerate low energy states, with the ultimate
ground state and low energy physics being sensitive to subtle
effects that act within this manifold. Due to this degeneracy, the
relevant energy scales are then much smaller than in conventional
unfrustrated systems and thus may arise from a wider variety of
sources. The competition between the small interactions can lead to a
rich set of phases, from unconventional magnetically ordered states or
even novel non-magnetic phases such as classical or quantum spin
liquids \cite{balents-2010-spin-liquid,gingras-mcclarty-2014-quantum,
springer-2011-frustrated}

\begin{figure}[htp]
  \centering
  \begin{overpic}[width=0.66\columnwidth]
    {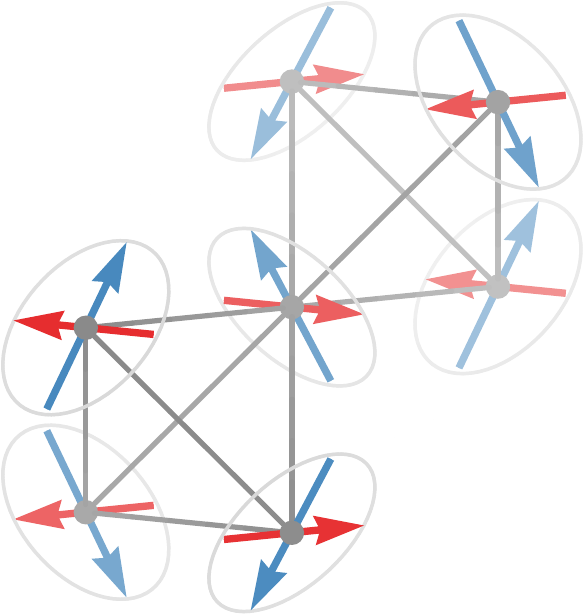}    
    \put(17,65){\textcolor{cblue}{\Large$\psi_2$}}
    \put(63,13){\textcolor{cred}{\Large$\psi_3$}}
    \put(80,33){\textcolor{clgray}{\Large$\Gamma_5$}}
    \put(28,91){\textcolor{cgray}{\Large Er\tsup{3+}}}
  \end{overpic}
  \caption{\label{fig:gamma5} 
    In \eto{} the Er\tsup{3+} ions form a pyrochlore lattice of
    corner-shared tetrahedra.The circles illustrate the allowed
    orientations in the ground state $\Gamma_5$ manifold, with the
    $\psi_2$ state (blue) and the $\psi_3$ state (red)
    indicated. The $\psi_2$ state is observed experimentally below
    $1.2\K$.
  }
\end{figure}
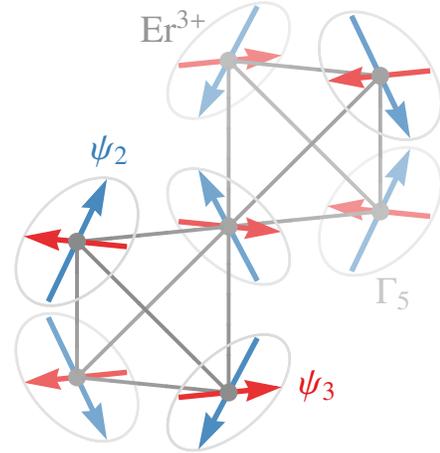

A particularly interesting class of degeneracy lifting mechanisms is
\emph{order by disorder}.  In the seminal incarnations of
\citet{villain-1980-order} and \citet{shender-1982-garnets}, thermal
or quantum fluctuations respectively select the states from the
degenerate manifold that have the largest space to fluctuate. This
perspective naturally leads to a generalization of the idea of order
by disorder, i.e. the selection from a degenerate manifold via \emph{any}
set of fluctuation corrections to the energy.  This mechanism of
fluctuation induced order is not only a concept within the purview of
frustrated magnetism, or even condensed matter physics taken more
broadly, but even appears in high-energy physics in the guise of the
Coleman-Weinberg mechanism \cite{coleman-weinberg-1973-radiative}.
This stabilization of order by quantum fluctuations, or \emph{order by
quantum disorder} \cite{shender-1982-garnets,
henley-1989-ordering,chubukov-1992-order}, has an even longer history.
While largely unnoticed, one of the earliest examples was put forth by
\citet{tessman-1954-magnetic} where quantum zero-point fluctuations
select the magnetization direction in a dipolar ferromagnet
\cite{van-1937-anisotropy,kittel-1951-dipolar}.

More formally, we consider a system whose ordering is described by
some order parameter $m$. The natural low-energy description can then
be framed in terms of the effective action $\Gamma[m]$.  In each order
by disorder scenario, one starts from an artificial limit where there
is an accidental degeneracy; that is the effective action at this
point, $\Gamma_0[m]$, has an accidental symmetry. For order by thermal
disorder this limit is $T \rightarrow 0$, while for order by
quantum-disorder this is the classical limit where the spin becomes
large, or $1/S \rightarrow
0$. Adding such perturbations moves the theory away from this artificial limit,
adding a term $\delta \Gamma [m]$ to the effective action as
$\Gamma[m] = \Gamma_0[m] + \delta \Gamma[m]$.  A distinction is
usually drawn between energetic and fluctuation-driven contributions
to selection, i.e. between zeroth order and higher order perturbative
corrections to $\delta \Gamma[m]$, though this can be somewhat
ambiguous.  Generically, $\delta \Gamma[m]$ breaks the accidental
symmetry, lifting the degeneracy present in $\Gamma_0[m]$.  In real
systems such a $\delta \Gamma[m]$ correction should \emph{always} be
present, so long as the symmetry is not exact, with each order by
disorder mechanism providing its own contribution to $\delta
\Gamma[m]$.  The identification of an order by disorder mechanism then
relies not only on a proximity to the idealized $\Gamma_0[m]$ limit,
but also on the contribution of that mechanism to $\delta \Gamma[m]$ being
dominant over all other sources of $\delta \Gamma[m]$ corrections.
Beyond having just a single dominant order by disorder channel, these
fluctuation selection mechanisms could thus cooperate and select the
same state, or perhaps compete and prefer different states
\cite{henley-1989-ordering,villain-1980-order}. Competing order by
disorder can lead to exotic multiple-step phase transitions, an
unusual sensitivity to perturbations or a host of other interesting
properties.

While it is easy for deviations from the initial idealized limit to spoil
an order by disorder selection mechanism, there are many systems
where the simplest energetic selection mechanisms can be forbidden
on fairly general grounds. Indeed, some of the early examples
\cite{tessman-1954-magnetic,shender-1982-garnets} focus on selection
of the moment direction in simple ferromagnets and antiferromagnets
with high symmetry. Classically, the presence of three-, four- or
six-fold rotations axes can forbid the selection of a moment
orientation purely from exchange anisotropy. In low-spin systems
without sufficient single-ion anisotropy, the leading anisotropy then
must come from effective multi-spin interactions. These can be
generated at the level of a Landau-Ginzburg theory through the order by
disorder mechanisms discussed above, be it thermal, quantum or
otherwise. To find a clear demonstration 
of this physics the challenge is then to find a material that
is sufficiently close to such a limit with accidental ground state
degeneracy. Theoretical and experimental control are \emph{essential}:
one must be able to characterize and quantify deviations
from the idealized theoretical limit with the degeneracy
while experimentally, one must be able to
validate the theoretical model and approach the limiting regime
closely, removing any extrinsic complications.

The class of rare-earth pyrochlore magnets \abo{$R$}{$M$} are promising
toward this goal \cite{gingras-gardner-2010-oxides,
gingras-mcclarty-2014-quantum}. Aside from having the requisite
high-symmetry, many are well described by an effective
spin-1/2 model \cite{ross-savary-2011-quantum-ice,
onoda-2011-quantum-fluctuations, savary-balents-2012-order} and thus
lack any on-site anisotropy.  At the nearest-neighbor level, there are
four symmetry allowed exchanges \cite{curnoe-2007-quantum-spin} leading to a diverse phase diagram
that includes a variety of magnetic and non-magnetic phases
\cite{benton-shannon-2013-edge,savary-balents-2013-ice}.  Included in
this phase diagram is a large region where the effective spins order
uniformly within their local XY planes \cite{wong-2013-generic-pyrochlore}.
This so-called $\Gamma_5$
manifold (see Fig. \ref{fig:gamma5}) can be thought of as a local ferromagnet and enjoys the 
degeneracy protection \cite{savary-balents-2012-order}
discussed above.  Of these rare-earth
pyrochlores, we will focus on the titanate \eto{} which is close to
ideal in each of these regards, having long been proposed as a prime
example of order by disorder \cite{champion-harris-2003}.
Experimentally, the thermodynamic properties
of this material are uncontroversial and have been
well-characterized.  The Er$^{3+}$ free ion hosts a $J=15/2$ manifold
that is quenched by the crystal field into an effective spin-1/2
degree of freedom with a strong planar (XY) anisotropy
\cite{champion-harris-2003}. The low-lying crystal field energy levels have been
measured to fair precision \cite{champion-harris-2003} and match well with theoretical models
\cite{malkin-2010-crystal-field,bertin-2012-crystal}.  Ordering occurs
near $T_{\rm N} \sim 1.2\K$ into the so-called $\psi_2$ state (see Fig. \ref{fig:gamma5}), a
non-coplanar antiferromagnet \cite{champion-harris-2003,poole-wills-2007}. The magnetic
excitations above this state are sharp and have been
measured across many cuts in reciprocal space via inelastic neutron
scattering \cite{ruff-gaulin-2008,savary-balents-2012-order,petit-gingras-2014-order}.  
A nearly gapless mode is observed \cite{ruff-gaulin-2008,
petit-gingras-2014-order,ross-gaulin-2014-order}, suggesting a quasi-degenerate set of ground
states, as needed for order by disorder.
Further, crystals of \eto{} can be made very clean, and disorder can be
re-introduced in a controlled manner via depletion or stuffing
\cite{gaulin-cava-2015-stuffing}.  The appropriate theoretical model
of the effective spin-1/2 degrees of freedom has been tightly
constrained by matching to the experimental data, through fitting of
the inelastic neutron scattering spectrum in both zero and finite
magnetic fields
\cite{savary-balents-2012-order,petit-gingras-2014-order}.
Altogether, this model can reproduce nearly all of the current
experimental data at a qualitative, and sometimes even a quantitative
level \cite{oitmaa-2013-phase}.

Classically, this effective spin-1/2 model has an accidental
degeneracy, the $\Gamma_5$ manifold mentioned above, consistent with
the soft-mode seen experimentally. Similarly, conventional mean-field theory
shows the same accidental degeneracy in the Landau-Ginzburg theory near $T_{\rm N}$ 
\cite{javanparast-gingras-2015-order}.
This classical degeneracy remains
unbroken for \emph{any} symmetry allowed, two-spin interactions
between the spins, to arbitrary distance \cite{savary-balents-2012-order}. 
Several proposals on how to lift
this degeneracy have been put forth; these include order by thermal disorder
(for $T \rightarrow 0^+$ or $T \rightarrow T_c^-$) 
\cite{champion-harris-2003,mcclarty-stasiak-gingras-2014-order,
zhitomirsky-moessner-2012-order,zhitomirsky-holdsworth-2014-nature,javanparast-gingras-2015-order}, order by quantum
disorder \cite{zhitomirsky-moessner-2012-order,savary-balents-2012-order,
mcclarty-stasiak-gingras-2014-order,wong-2013-generic-pyrochlore} 
or order by structural disorder
\cite{maryasin-zhitomirsky-2014-order,mcclarty-2015-order-dilution}.
A case has been made for order by quantum disorder
\cite{zhitomirsky-moessner-2012-order,savary-balents-2012-order} as
the operant mechanism in lifting the degeneracy of the classical
ground state manifold in \eto{}, with order by thermal disorder
\cite{oitmaa-2013-phase} cooperating at the ordering 
temperature $T_{\rm N}$.  A key prediction of the order by quantum disorder
 proposal is a gap in the
excitation spectrum of order $\sim 20\mueV$, comparable to the $\sim
40-45\mueV$ gap that was later observed experimentally
\cite{ross-gaulin-2014-order,petit-gingras-2014-order}.  One
\emph{implicit} limit that has been taken  in all these studies is to assume that an
effective spin-1/2 model with only two-spin interactions is itself
sufficient. This can justified if one can take the crystal field
energy scale $\eta$ to be much larger than the exchange scale. Taking
$\eta$ to be finite then allows for corrections to the effective
spin-1/2 model through virtual crystal field excitations, yet another 
channel contributing to $\delta \Gamma[m]$ as defined above.  While these
terms \cite{mcclarty-2009-energetic,petit-gingras-2014-order} 
have been argued to be extremely small \cite{savary-balents-2012-order},
the complexity of the multipolar interactions between rare-earth ions
\cite{elliott-1968-orbital, santini-2009-multipolar,
iwahara-2015-exchange, rau-2015-quantum-ice} and the subtleties arising
from the combinatorics of
high-order perturbation theory demand a more careful accounting of
these effects.

In this article, we aim to carefully address the effects of virtual crystal
field fluctuations \cite{mcclarty-2009-energetic,petit-gingras-2014-order}
on the ground state selection in \eto{}. We find that such an
\emph{order by virtual crystal field fluctuations} is not only
significant, but is naturally comparable with the effects from order by quantum
disorder. When mapped into the effective spin-1/2 description via
strong coupling perturbation theory, these fluctuations manifest as
multi-spin interactions, with a complementary interpretation as
classical energetic selection. We argue that the complex multipolar
interactions that exist in the full $J=15/2$ manifold further enhance these
fluctuations, compared to the case of only bilinear interactions
between the angular momenta $\vec{\mJ}$, and render them very relevant in describing \eto{}.  
For definiteness, we consider a bilinear-biquadratic exchange model 
of the $J=15/2$ moments as a minimal model of this
physics. We analyze the effects of virtual crystal field
fluctuations using a combination of mean-field theory (MFT) and
random-phase approximation (RPA) calculations.  We show that these
MFT+RPA calculations are qualitatively consistent with a treatment of
the multi-spin interactions via strong-coupling perturbation theory,
and thus capture the effects of the virtual crystal field excitations.  Generically,
we find that this mechanism is cooperative with thermal and quantum
selection, selecting the $\psi_2$ state as is seen experimentally.  From
these calculations we find that when the bilinear and biquadratic
couplings are comparable this selection mechanism can be of similar magnitude or
even \emph{dominate} over the effects of order by quantum disorder 
\cite{savary-balents-2012-order,zhitomirsky-moessner-2012-order}.  
We thus conclude that \eto{} may not ultimately be the long sought definite
case for order by quantum disorder controlling ground state selection.
More broadly, we have identified an important fluctuation channel in
rare-earth magnets, that highlights some of the pitfalls
that can may be hidden in the effective spin-1/2 theory . 
Knowledge of the leading mechanisms for ground
state selection is important in the search for candidates that realize
order by disorder.  While cooperative in \eto{}, similar
considerations may yield competing mechanisms for state selection in other
materials, with associated multiple phase transitions or other
exotic behaviors. Finally, we discuss how these considerations may apply to
other materials in this family, such as the pyrochlore
magnets \abo{Er}{$M$} or \abo{Yb}{$M$}.

The paper is organized as follows: we begin
Sec. \ref{sec:effective-models} by reviewing the low-energy effective
spin-1/2 model for \eto{}. We then show in
Sec. \ref{sec:multipolar-interactions} and
Sec. \ref{sec:strong-coupling} how this effective model arises from
underlying multipolar interactions between the $J=15/2$ manifolds of
the Er\tsup{3+} ions, and discuss the generation of multi-spin
interactions by virtual crystal field fluctuations.  We build on this
in Sec. \ref{sec:obvcef}, showing how these multi-spin interactions
lead to ground state selection and order by virtual crystal field
fluctuations.  In Sec. \ref{sec:classical}, we show how this selection
proceeds and how it is related to the gap in the spin-wave spectrum,
presenting estimates for their magnitude in
Sec. \ref{sec:scaling}. Next, in Sec. \ref{sec:biquadratic}, we
perform an explicit calculation of these quantities for two models
of multipolar interactions in \eto{}. As discussed in
Sec. \ref{sec:super-exchange} and Sec. \ref{sec:eqq}, these models
include both bilinear and biquadratic exchange, and are solved with
the RPA approximation as outlined in Sec. \ref{sec:rpa}. The results
for ground state selection and the excitation spectrum are presented
in Sec. \ref{sec:results}. The application of these results to \eto{}
and how to disentangle different order by disorder scenarios are
discussed in Sec. \ref{sec:discussion}. Finally, we comment in
Sec. \ref{sec:conclusion} on possible implications for
other materials.

\section{Effective models}
\label{sec:effective-models}
Before delving into the detailed modeling of \eto{}, we first summarize 
the arguments that lead to the effective spin-1/2 model that 
has been used in prior works \cite{savary-balents-2012-order,oitmaa-2013-phase}.
The ground state of the free Er\tsup{3+} ion consists of the
${}^4I_{15/2}$ manifold of the $4f^{11}$ electronic configuration which carries a
$J=15/2$ moment.  In the pyrochlore structure of \eto{} (space group
$Fd\cb{3}m$), the Er\textsuperscript{3+} ion lies at a site with
$D_{3d}$ symmetry. The crystal field lifts the degeneracy of the
${}^4I_{15/2}$ manifold, splitting it into a set of effective spin-1/2 $\Gamma_4$ doublets
and dipolar-octupolar $\Gamma_5 \oplus \Gamma_6 \equiv \Gamma_{56}$ doublets
\cite{huang-chen-2014-octupolar}.  The ground
doublet is an effective spin-1/2 degree of freedom of type $\Gamma_4$ with a gap of
$\cefS \sim 6\meV$ ($\sim 70\K$) to the first excited state \cite{champion-harris-2003}. 
 Since this energy scale
is large relative to the exchanges, which are expected to be $\sim
0.1\K-1\K$ \cite{savary-balents-2012-order,petit-gingras-2014-order}, 
it is justified to consider an effective spin-1/2 model,
down-folding the full set of crystal field levels into the ground doublet
defined as $\ket{\pm}$.
In the coarsest approximation, one simply projects the $J=15/2$
moment $\vec{\mJ}_i$ at site $\vec{r}_i$ into this manifold
\begin{equation}
  P \vec{\mJ}_i P = \lambda_{\pm}\left(\mS^x_i \vhat{x}  +\mS^y_i\vhat{y}  \right)+
  \lambda_z \mS^z_i \vhat{z} \equiv \mat{\lambda} \vec{\mS}_i,
\end{equation} 
where $\vec{S}_i$ is the effective spin-1/2 operator and
$\mat{\lambda} \equiv {\rm
diag}(\lambda_{\pm},\lambda_{\pm},\lambda_z)$ with $\lambda_{\pm}\sim
6$ and $\lambda_z\sim 2$ depends on the details of the crystal field
parameters through the spectral composition of the ground
doublet. These $\lambda$-factors are related to the $g$-factors which
characterize the response to an applied magnetic field via
$\lambda_{\mu} = g_{\mu}/g_J$ where $\mu=\pm,z$ and $g_J=6/5$ is the
Land\'e factor for Er$^{3+}$.  Both the $\vec{J}_i$ and $\vec{S}_i$
operators are defined in the local basis $(\vhat{x}_i,
\vhat{y}_i,\vhat{z}_i)$ defined with respect to the high-symmetry
directions at each site, as defined in Appendix \ref{basis}. In this
local picture, the $\Gamma_5$ manifold corresponds to ferromagnetic
ordering in the local XY plane, with $\psi_2$ defined as the
moments ordered in the local $\vhat{x}$ direction (see Fig. \ref{fig:gamma5}). 
The remaining states can be obtained from $\psi_2$ by a rotation of
of all the spins about their local $[111]$ direction.

As discussed in the Introduction, the most detailed information on the
interactions with the Er$^{3+}$ ions comes from fitting such an
effective spin-1/2 model to the results of inelastic neutron
scattering experiments
\cite{savary-balents-2012-order,petit-gingras-2014-order}.  It was
found that a model with only nearest-neighbor anisotropic exchange
between the effective spin-1/2 moments provides a fairly good
description of the observed spectrum. By symmetry, 
the effective spin-1/2 Hamiltonian takes the form
\cite{curnoe-2007-quantum-spin,savary-balents-2012-order}
\begin{align}
  \label{eq:s-model}
  H_{\rm eff} &= 
\sum_{\avg{ij}} \trp{\vec{\mS}}_i \mat{\exJ}_{ij} \vec{\mS}_j\nonumber \\
&\equiv \sum_{\avg{ij}} \left\{  
  \exJzz \mS^z_i \mS^z_j - \exJpm\left(\mS^+_i\mS^-_j+\mS^-_i\mS^+_j\right)+\right. \\
& \exJpp \left(\gamma_{ij} \mS^+_i\mS^+_j+{\rm h.c}\right)+  
 \exJzp \left. \left(\nonumber
    \zeta_{ij} \left[ \mS^z_i \mS^+_j+ \mS^+_i \mS^z_j \right]+ {\rm h.c}\right) \right\}.
\end{align}
Details of the complex form factors $\gamma_{ij}$ and $\zeta_{ij}$ are
given in Appendix \ref{basis}. From experimental fitting
\cite{savary-balents-2012-order,petit-gingras-2014-order}, the $\exJ_{ij}$ couplings are
of order $10^{-2}\meV$ with $\exJ_{\pm}$ and $\exJ_{\pm\pm}$ being the
largest. A representative example, from
Ref. [\onlinecite{savary-balents-2012-order}], is
\begin{align}
  \label{eq:exp-params}
  \exJ_{zz}     &= -2.5 \times 10^{-2} \meV,   \nonumber &
  \exJ_{\pm}    &= +6.5 \times 10^{-2} \meV,  \\
  \exJ_{\pm\pm} &= +4.2  \times 10^{-2} \meV,    &
  \exJ_{z\pm}   &= -0.88 \times 10^{-2} \meV.
\end{align}
Further details on other proposed sets of effective spin-1/2 exchanges
are found in Appendix \ref{exchanges}.  Longer range couplings, such
as second or third neighbor exchanges or dipolar interactions are
generally expected to be present, but have not been found to be necessary to
reproduce the features that are theoretically accessible within
current experimental uncertainties. 
While this model has been found to be
fairly successful, we show, as
outlined in the Introduction, that higher corrections that go
beyond the coarse approximation encapsulated in Eq. (\ref{eq:s-model})
will prove to be significant. To include such corrections, we must
build a model for \eto{} starting from the atomic physics and
progressing towards the low-energy effective Hamiltonian.

\subsection{Multipolar interactions}
\label{sec:multipolar-interactions}
The structure of the model (\ref{eq:s-model}) was essentially
fixed by symmetry and the restriction to only two-spin (bilinear) interactions between
the effective spin-1/2 degrees of freedom.
Information on the microscopic interactions between the $\vec{\mJ}_i$ moments or the $4f$
electrons themselves is essentially lost through the projection into the
ground doublets; compressed into the four
exchange parameters $\exJ_{zz}$, $\exJ_{\pm}$, $\exJ_{\pm\pm}$ and
$\exJ_{z\pm}$. To go beyond this projection into the ground
doublets, we move up in energy and consider a model of the multipolar interactions between the $J=15/2$
moments. As above, we consider only interactions between
nearest-neighbor sites. This can be partially justified as, aside from
the long-range dipolar interactions, one expects multipolar couplings
to arise form short-range super-exchange type processes
\cite{elliott-1968-orbital, santini-2009-multipolar}.

In contrast to the effective spin-1/2 model of Eq. (\ref{eq:s-model}),
we do \emph{not} expect these more microscopic interactions to have only a bilinear
form. Unlike the effective spin-1/2 moments, the $J=15/2$ manifolds of
the Er\tsup{3+} ions can support many higher-order multipoles. These
multipoles can be classified into \emph{ranks}; the familiar dipole
operators $\sim \mJ^\alpha$ are rank-1, quadrupole operators $\sim
J^{\alpha} J^{\beta}$ are rank-2 and so forth. 
As $J=15/2$, such
multipoles can be constructed up to and including rank-15. To proceed
in a systematic fashion we introduce a basis for these multipole operators
that transforms in the same way as the spherical harmonics. As for 
the spherical harmonics, for a rank-$K$ multipole we have $2K+1$ operators
indexed by $Q=-K,-K+1, \ldots +K-1,+K$. More explicitly, we define 
the set of rank-$K$ multipoles $O_{KQ}(\vec{\mJ})$ 
\begin{equation}
  \label{multipole}
\brap{J,M} O_{KQ}(\vec{\mJ}) \ketp{J,M'} \equiv 
\sqrt{\frac{2K+1}{2J+1}}\braketp{J,M;K,-Q}{J,M'},
\end{equation}
where $\braketp{J,M;K,-Q}{J,M'}$ is a Clebsch-Gordan coefficient
and $\ket{J,M}$ are eigenstates of $\vec{\mJ} \cdot \vec{\mJ}$ and $J^z$.
These operators have been normalized so that
$\tr{[\h{O_{KQ}(\vec{\mJ})} O^{}_{KQ}(\vec{\mJ})]} = 1$, so the
different ranks can be compared on equal footing without worrying about large matrix
element factors that arise when acting on the $\ket{J,M}$ states.  The
rank-1 vector operators are then simply a reformulation of the dipole
moment $\vec{\mJ}_i$
\begin{align}
  \label{eq:vectortensor}
  O_{1,0}(\vec{\mJ}) &= \frac{1}{2\sqrt{85}} \mJ^z,
&  O_{1,\pm 1}(\vec{\mJ}) &=  \pm \frac{1}{2\sqrt{85}}\left(
                            \frac{J^x \pm i J^{y}}{\sqrt{2}}\right),
\end{align}
where the factor of $1/(2\sqrt{85})$ enforces the trace normalization.
We note that these operators are similar to, but not identical to the
Stevens' operator equivalents \cite{stevens-1952-matrix} used to
define the crystal field potential.

With these degrees of freedom in hand, we can write down a model
of their interactions. Aside from the crystal field potential, because of the
weakness of super-exchange processes in $4f$ insulators \cite{santini-2009-multipolar} 
we expect them to be predominantly pair-wise, generically giving the model
\begin{equation}
  \label{eq:j-model}
  \sum_{\avg{ij}} \sum_{KQ} \sum_{K'Q'} O_{KQ}(\vec{\mJ}_i) \exM^{KQ,K'Q'}_{ij} O_{K'Q'}(\vec{\mJ}_j) +
  \sum_i \cfV(\vec{\mJ}_i),
\end{equation}
where $\cfV(\vec{\mJ})$ is the crystal field potential,
$O_{KQ}(\vec{\mJ}_i)$ is a multipole of rank-$K$ and $\mat{\exM}$ are
the multipolar coupling constants.  The form and parameters we use for
$\cfV(\vec{\mJ})$ are discussed in Appendix \ref{crystal-field}.
Processes such as super-exchange can only generate interactions up to
including rank-7 multipoles \cite{elliott-1968-orbital,
santini-2009-multipolar, iwahara-2015-exchange, rau-2015-quantum-ice}.
Given that the anisotropy in the fitted exchanges in \eto{} deviates
strongly from that expected from magnetostatic dipolar interactions,
we expect the super-exchange scale to be significant, and thus a wide
range of multipolar ranks to be present in the interactions of the
$J=15/2$ model given by Eq. (\ref{eq:j-model}). There are an enormous
number of independent couplings embedded in the matrix $\mat{\exM}$,
given that $K,K' \leq 7$ and $|Q| \leq K$, $|Q'|\leq K'$. Even
accounting for symmetry leaves hundreds of possible couplings.
Thankfully, we have some semblance of a separation of scales via the
crystal field potential, which at $\cefS \sim 6 \meV$ is roughly two orders
of magnitude larger than the expected scale for $\exM$.

We now have all the pieces to improve on the effective model of
Eq. (\ref{eq:s-model}), which can, in principle, be obtained for a given
$\mat{\exM}$ by a bare projection of Eq. (\ref{eq:j-model}) into the ground
doublets. In the following section we sketch a derivation of the low
energy effective model that goes beyond simply the bilinear
interactions defined in Eq. (\ref{eq:s-model}). 
We do this leaving the multipolar interactions somewhat
arbitrary, relegating the detailed discussion of their form to
Sec. \ref{sec:biquadratic}.

\subsection{Strong coupling perturbation theory}
\label{sec:strong-coupling}
To derive the low-energy effective model, we carry out strong coupling perturbation
theory in the crystal field potential, perturbing with the multipolar
couplings. Aside from the simple model of Eq. (\ref{eq:s-model}), we
also consider the higher order corrections that have been
ignored in previous studies.  More concretely, we use the
Rayleigh-Schr\"odinger perturbation theory as presented in
\citet{lindgren-1974-rayleigh}.  We decompose the full
Hamiltonian, $H$, as $H = \eta H_0 +V$
where
\begin{subequations}
\begin{align}
  H_0 &= \sum_i \cfV(\vec{\mJ}_i),\\
  V  &=  \sum_{\avg{ij}} \sum_{KQ} \sum_{K'Q'} 
        O_{KQ}(\vec{\mJ}_i) \exM^{KQ,K'Q'}_{ij} O_{K'Q'}(\vec{\mJ}_j).
\end{align}
\end{subequations}
To control
the passage from the $J=15/2$ model to the effective spin-1/2, we have
added a dimensionless rescaling parameter $\eta$ to the crystal field. The choice $\eta=1$
corresponds to the physical, experimentally fitted crystal field Hamiltonian
\cite{bertin-2012-crystal}, with larger values suppressing the effects
(i.e. admixing) of the higher lying crystal field doublets 
in the low-energy sector composed of the ground doublets.
The bare Hamiltonian $\eta H_0$ is diagonalized using single-ion
crystal field states at each site.  To proceed with the perturbation theory, we
define a projector into the ground state manifold, $P = \sum_{E_n =
E_0} \ket{n}\bra{n}$, along with a resolvent operator
\begin{equation}
 R = \sum_{E_n \neq E_0} \frac{\ket{n}\bra{n}}{E_0 -E_n},
\end{equation}
where $E_n$ and $\ket{n}$ are the eigenvalues and eigenstates of
$H_0$.  This perturbation theory becomes formally exact in the limit
$\eta \rightarrow \infty$. The expansion of the effective Hamiltonian
is given in \citet{lindgren-1974-rayleigh} as $H_{\rm eff} = \sum_{n}
\eta^{-n+1} \Heff{n}$.  Here one considers a wave-operator, $\Omega$,
related to the effective Hamiltonian as $H_{\rm eff} = PV\Omega$ with
the analogous expansion $\Omega = \sum_n \eta^{-n+1}\Omega_n$. These
are defined via the relation 
\begin{equation}
  \label{recurse} \Omega_n = RV \Omega_{n-1} -R \sum_{m=1}^{n-1}
\Omega_{n-m} V \Omega_{m-1}.
\end{equation} Starting with $\Omega_1 = P$, the higher order
$\Omega_n$, and thus $H_{\rm eff,n}$, can be computed recursively.
The first three orders are given by
\begin{subequations}
\label{eff-exprs}
\begin{eqnarray}
  \Heff{1} &=& P V P, \\
  \Heff{2} &=& P V R V P, \\
  \Heff{3} &=& P VRVRVP  - 
  \frac{1}{2}
  \left\{(P V R^2 VP),P VP\right\},
\end{eqnarray}
\end{subequations}
where $\{\cdot,\cdot \}$ is the anti-commutator
\footnote{The higher order terms $\Heff{3}$ and $\Heff{5}$ have been
symmetrized via $\Heff{n} \rightarrow (\Heff{n} + \h{\Heff{n}})/2$ to
render the effective Hamiltonian Hermitian. This procedure can be
justified in that there exists a transformation that maps the
unsymmetrized form into the symmetrized form, as discussed by
\citet{des-1960-extension}.
}. 
Due to their complexity, we will not give the expressions for $H_{\rm
eff,4}$ and $H_{\rm eff,5}$, but they can be straightforwardly
computed from Eq. (\ref{recurse}).

The leading term, $\Heff{1}$, is simply the projection of the
multipolar interactions, $V$, into the ground doublets.  This can be
accomplished by projecting the multipoles individually
\begin{equation}
  \label{eq:proj}
  P O_{KQ}(\vec{\mJ}_i) P = 
  \begin{cases}
     \vec{v}_{KQ} \cdot \vec{\mS}_i& \text{if } K \text{ odd}\\
    \text{const.} & \text{if } K \text{ even}
  \end{cases},
\end{equation}
then substituting them back into $V$ in Eq. (\ref{eq:j-model}).  
The vectors $\vec{v}_{KQ}$ are in general complex and define the mapping
into the effective spin-1/2, depending only on the spectral composition of the
ground doublet.  At first order in perturbation theory, the even rank terms lead only to a
constant shift of the energy and can be ignored.  As such, we see that
the even-rank multipolar interactions are \emph{completely
unconstrained} by the fitted effective-spin 1/2 exchanges, $\exJ_{ij}$.  Combined together,
the odd terms give the interactions between the effective spin-1/2 operators
$\vec{\mS}_i$ as
\begin{equation}
  \label{heff1}
 H_{\rm eff,1} = \sum_{\avg{ij}} \trp{\vec{\mS}}_i \left(
\sum_{KQ}\sum_{K'Q'} \vec{v}_{KQ} \trp{\vec{v}}_{K'Q'} \exM^{KQ,K'Q'}_{ij}
  \right) \vec{\mS}_j,
\end{equation}
This leads to the symmetry allowed nearest-neighbor model as
given in Eq. (\ref{eq:s-model}), with only two-spin interactions, where
the tremendous complexity of $\mat{\exM}$ is embedded in the four symmetry
allowed exchanges $\exJ_{zz}$, $\exJ_{\pm}$, $\exJ_{\pm\pm}$ and $\exJ_{z\pm}$.

At higher order, further corrections will be generated through virtual
crystal field fluctuations \cite{molavian-2007-dynamically} involving
the excited single-ion states through the resolvent operators that appear
in Eq. (\ref{eff-exprs}). The leading corrections that do not simply
renormalize the bilinear couplings are four- and six-spin interactions \cite{
wong-2013-generic-pyrochlore,javanparast-gingras-2015-order}
which appear at third and fifth order in $\exM$. We thus write the effective
Hamiltonian at fifth order in three pieces, $H_{\rm eff} = H_2 + H_4 +
H_6$ containing the two-, four- and six-spin interactions. Explicitly,
we have
\begin{subequations}
\label{eff-ham}
\begin{align}
H_2 &=
 \sum_{{i_1 i_2}} \sum_{\alpha_1 \alpha_2} \exJ^{\alpha_1\alpha_2}_{i_1i_2} \mS^{\alpha_1}_{i_1} {\mS}^{\alpha_2}_{i_2}, \\
H_4 &= \sum_{{i_1 \cdots i_4}} \sum_{\alpha_1 \cdots \alpha_4}
\exK^{\alpha_1 \cdots \alpha_4}_{i_1 \cdots i_4}
\mS^{\alpha_1}_{i_1} \mS^{\alpha_2}_{i_2} \mS^{\alpha_3}_{i_3}  \mS^{\alpha_4}_{i_4}, \\
H_6 &= \sum_{{i_1 \cdots i_6}} \sum_{\alpha_1 \cdots \alpha_6}
\exL^{\alpha_1 \cdots \alpha_6}_{i_1 \cdots i_6}
\mS^{\alpha_1}_{i_1} \mS^{\alpha_2}_{i_2} \mS^{\alpha_3}_{i_3} 
\mS^{\alpha_4}_{i_4} \mS^{\alpha_5}_{i_5} \mS^{\alpha_6}_{i_6}.
\end{align}
\end{subequations}
The lowest order part of the bilinear couplings $\exJ$ in Eq. (\ref{eff-ham}) are the nearest
neighbor projection of the multipolar interactions at order $\exM$, as
given in Eq. (\ref{heff1}). Higher corrections at orders $\exM^2/\cefX{}$
up to $\exM^5/\cefX{4}$ renormalize bilinear couplings between
the effective spin-1/2 operators, as well as
inducing second, third and further neighbor exchanges where $\cefS$ defines the
scale of crystal field when $\eta=1$. The nearest-neighbor 
part will be taken to match the experimentally determined
values of Eq. (\ref{eq:exp-params}). The subleading further neighbor
exchanges do not lift the $\Gamma_5$ degeneracy on their own
\cite{savary-balents-2012-order} and serve only to renormalize the
bilinear exchange scale. While this will modify the spin-wave
dispersion and other features of model, it will not affect the
ground state selection or excitation gap. Since this is our focus and the
question of foremost interest, we ignore these
terms.

The four-spin interactions, $\exK$, appear first at order
$\exM^3/\cefX{2}$ and receive subleading corrections from fourth- and
fifth-order as $\exM^4/\cefX{3}$ and $\exM^5/\cefX{4}$.  We keep only the
leading terms, at order $\exM^3/\cefX{2}$, where the four sites $i_1,
\cdots, i_4$ take the form of trees built from three nearest-neighbor
bonds of the lattice.  For example, a tree for three bonds can involve
four distinct sites and thus contribute to the four-spin operators at
order $\exM^3/\cefX{2}$. Contributions with three bonds that have a loop
involve at most two or three spins and can not contribute to a
four-spin term.  Similarly, for the six-spin interactions, $\exL$, the
leading contribution comes in at order $\exM^5/\cefX{4}$ with the sites
$i_1,\cdots,i_6$ forming trees built from five nearest-neighbor bonds.
A large number of possible anisotropic couplings are allowed through
the dependence on the spin indices in both the four- and six-spin
interaction terms.

\section{Order by virtual crystal field fluctuations}
\label{sec:obvcef}
Having understood how higher spin interactions are induced by virtual
crystal field excitations, we next need to understand how the four-
and six-spin interactions affect the ground state selection and the
energy gap. To do this, we construct a variational product state for
each local ferromagnetic ordering direction $\ket{\vhat{n}}$ as
\begin{equation}
\label{product}
\ket{\vhat{n}}  \equiv 
\prod^N_{i=1} \left[
\cos\frac{\theta}{2}\ket{\up}_i + 
e^{i\phi} \sin\frac{\theta}{2} \ket{\down}_i\right],
\end{equation}
where $\ket{\up}_i$, $\ket{\down}_i$ are effective spin-1/2 states of
the low-energy effective model and $(\theta,\phi)$ are the spherical
angles for the direction $\vhat{n}$.  By construction, the expectation
value of the effective spin-1/2 is oriented in the direction specified
by $\theta$ and $\phi$
\begin{align}
  \label{product-ex}
  \exc{\vhat{n}}{\vec{\mS}_i}{\vhat{n}} 
&= \frac{1}{2} \left[
    \sin\theta \left(\cos\phi \vhat{x} +\sin\phi \vhat{y}\right) + \cos\theta \vhat{z}
  \right] = \frac{1}{2} \vhat{n}.
\end{align}
The $\Gamma_5$ manifold is then the set of states with $\theta=\pi/2$;
$\psi_2$ corresponding to $\phi=n\pi/3$ and $\psi_3$ to $\phi=n\pi/3 +
\pi/6$ with $n=0,1,2,\ldots 5$.

When phrased in terms of the states in the low-energy effective model,
the state $\ket{\vhat{n}}$ contains no inter-site correlations.
Thus when viewed through the lens of the low-energy effective model,
the selection via multi-spin
interactions could be regarded as a purely \emph{classical} energetic effect. 
 While intuitive, such a perspective obscures some of
the key aspects of the physics.  A more complete viewpoint requires
consideration of what $\ket{\vhat{n}}$ corresponds to in terms of the physical
$\ket{\pm}_i$ crystal field ground doublets. At zeroth order one would
simply map the effective spin-1/2 states into the ground doublets
directly, replacing $\ket{\up}_i \rightarrow \ket{+}_i$ and
$\ket{\down}_i \rightarrow \ket{-}_i$ in Eq. (\ref{product}).  At
higher orders in perturbation theory the relationship between the
states of the low-energy effective theory and the underlying $J=15/2$
manifold becomes non-trivial. Following
\citet{lindgren-1974-rayleigh}, the $\ket{\vhat{n}}$ state maps to
\begin{equation}
  \ket{\vhat{n}}_{\Omega} \equiv \Omega \ket{\vhat{n}},
\end{equation}
where $\Omega$ is the wave-operator whose expansion was given in
Eq. (\ref{recurse}). The physical state $\ket{\vhat{n}}_{\Omega}$ is
\emph{not} a product state in the ground doublets $\ket{\pm}_i$.  The
wave-operator $\Omega$ has much of the same structure as the effective
Hamiltonian and is \emph{non-local} in the moment operators
$\vec{\mJ}_i$. This mapping encodes non-trivial correlations between
the sites, as well as virtual fluctuations into the higher crystal
field levels.  

These contributions to degeneracy lifting are thus induced through
virtual fluctuations into the higher crystal field levels of the
$J=15/2$ manifold that are built into the state itself. We
therefore refrain from referring to this ground state
selection as ``energetic'', and
thus call this mechanism \emph{order by virtual
crystal field fluctuations}. Most importantly, this is distinct from the correlations
built into the state in the order by quantum disorder scenario 
\cite{savary-balents-2012-order}; the
correlations induced here \emph{vanish} as the crystal field energy
scale becomes large, i.e. as $\eta \rightarrow \infty$.

Given this mapping, one may ask how the direction $\vhat{n}$ relates
to direction of the physical moments $\vec{\mJ}_i$. For the
$\psi_2$ ($\phi = n\pi/3$) and $\psi_3$ ($\phi = n\pi/3 + \pi/6$)
states of interest, this is partially constrained by their remnant 
symmetry.  Each $\psi_3$ state is preserved (locally)
by the $C_2$ rotation along the moment direction, while each $\psi_2$
state is preserved by a combination of a $C_2$ axis perpendicular to
the moment, combined with time-reversal. As the mapping $\Omega$ preserves
the underlying space-group symmetries of the model, these remnant symmetries
are also reflected in the expectations of the moments $\vec{\mJ}_i$. This then
implies that the local in-plane angle is preserved by $\Omega$ for the $\psi_2$
and $\psi_3$ states. We note that for $\psi_3$ the vanishing of the
component out of the local $[111]$ plane is also preserved
with the expectation $\vec{\mJ}_i$ remaining fixed in the local XY planes. 
For $\psi_2$, a finite
canting out of the XY plane \cite{javanparast-gingras-2015-order} is generically induced by this mapping, in addition to
an intrinsic generation that will be discussed in the following section.

\subsection{Selection and excitation gap}
\label{sec:classical}
We can now look at the selection of the $\psi_2$ or $\psi_3$ state and
the excitation gap that follows generically.  Using the
effective spin-1/2 language, we compute the
energy per spin as
\begin{equation}
  E(\vhat{n}) \equiv \exc{\vhat{n}}{H_{\rm eff}}{\vhat{n}} /N.
\end{equation}
Practically, one simply replaces each effective spin-1/2 operator in
$H_{\rm eff}$ with the classical spin vector in
Eq. (\ref{product-ex}), dividing by the total number of sites. The
energy per spin is then given by
\begin{align}
  \label{eq:cl-en}
  E(\vhat{n}) &= C_2 \cos^2 \theta
        +C_4 \cos\theta \sin^3\theta \cos 3\phi
        -C_6 \cos 6\phi,
\end{align}
where the overall form is dictated by symmetry constraints, as has
been discussed in Refs. [\onlinecite{wong-2013-generic-pyrochlore},
\onlinecite{javanparast-gingras-2015-order}].  The $C_n$
coefficients are determined by the exchanges in Eq. (\ref{eff-ham})
and are given by
\begin{subequations}
\begin{align}
C_2 &=
 \frac{1}{2} \sum_{{i_1 i_2}} \sum_{\alpha_1 \alpha_2} \exJ^{\alpha_1\alpha_2}_{i_1i_2}, \\
C_4 &= \frac{1}{2^4}\sum_{{i_1 \cdots i_4}} \sum_{\alpha_1 \cdots \alpha_4}
{\exK}^{\alpha_1 \cdots \alpha_4}_{i_1 \cdots i_4}, \\
C_6 &= \frac{1}{2^6}\sum_{{i_1 \cdots i_6}} \sum_{\alpha_1 \cdots \alpha_6}
{\exL}^{\alpha_1 \cdots \alpha_6}_{i_1 \cdots i_6}.
\end{align}
\end{subequations}
At zeroth order in $1/\eta$, the $C_2$ coefficient can be worked out from the bare
projected couplings with
\begin{equation}
  C_2 \approx 3 (\exJ_{zz}+2\exJ_{\pm}).
\end{equation}
The value of this coefficient does not depend too much on the choice of the exchange parameters;
using the parameters of Ref. [\onlinecite{savary-balents-2012-order}]
one has $C_2 \sim 0.3 \pm 0.1 \meV$, while from
Ref. [\onlinecite{petit-gingras-2014-order}] one has $C_2 \sim 0.3 \pm
0.05 \meV$.  The contributions from the $\exJ_{\pm\pm}$ and
$\exJ_{z\pm}$ terms cancel for local ferromagnetic states due to their
complex form factors $\gamma_{ij}$ and $\zeta_{ij}$ in Eq. (\ref{eq:s-model}).  
Such a cancellation is generic
\cite{savary-balents-2012-order}, as a by-product of lattice symmetries
further guaranteeing that further bilinear interactions of arbitrary range 
do not lift the degeneracy of the $\Gamma_5$ manifold.

To see which state is selected, we minimize the total energy per spin, $E(\vhat{n})$,
as a function of $\theta$ and $\phi$.  Given that the coefficient
$C_2$ is positive and leading for the parameters relevant to \eto{},
we expect predominantly in-plane order ($\theta \sim \pi/2$).  We thus can expand about
$\theta = \pi/2$ as
\begin{equation}
  E(\vhat{n}) \sim C_2 \left(\theta - \frac{\pi}{2}\right)^2 -
C_4 \left(\theta - \frac{\pi}{2}\right) \cos(3\phi) + \cdots
\end{equation}
where the omitted terms do not depend on $\theta$.  This gives a
minimum at
\begin{equation}
  \label{eq:mz}
  \theta_0 = \frac{\pi}{2} + \left(\frac{ C_4}{2C_2} \right) \cos(3\phi).
\end{equation}
Fixing $\theta = \theta_0$ and using Eq. (\ref{eq:cl-en}), the energy difference
\footnote{We note that contributions to the energy such as $\cos(6 n\phi)$ are
also allowed. Since each of these terms carries a factor $S^{6n}$, so
for $S=1/2$ we will neglect all but the $n=1$ case which has been
considered.} 
between the $\psi_2$ and $\psi_3$ states is then
\begin{equation}
\label{eq:endiff}
\delta E \equiv E({\psi}_3) - E({\psi}_2)= 2 \left(\frac{C^2_4}{8C_2} +  C_6\right).
\end{equation}
If $C_4^2 + 8C_2 C_6 > 0$, then the $\psi_2$ states are selected and
otherwise $\psi_3$ is selected.  The $\psi_2$ ($\phi=n\pi/3$) states are modified by a
small canting out of the local XY plane as indicated by $\theta_0 \neq \pi/2$ \cite{javanparast-gingras-2015-order} in Eq. (\ref{eq:mz}).  One
sees then that a non-zero $C_4$ \emph{always} favors the selection of
$\psi_2$ while a finite $C_6$ can select $\psi_2$ or $\psi_3$
depending on its sign. This can be seen in by the absence of canting
in the $\psi_3$ state where $\theta_0 = \pi/2$ for $\phi_0 = n\pi/3 + \pi/6$ in
Eq. (\ref{eq:mz}).  Thus $\psi_3$ cannot
gain any energy from the $C_4$ term and must be stabilized solely by
$C_6$.

Having identified how $\psi_2$ or $\psi_3$ are selected within the $\Gamma_5$ manifold,
we estimate the size of the spin-wave gap in the $\psi_2$ state.
Solving the classical equations of motion for a finite energy
$\vec{k}=0$ mode, the gap is given by \cite{jensen-1991-rare}
\begin{equation}
  \label{gap-one}
  \Delta =2 \sqrt{{{A}}_{\phi\phi}
{{A}}_{\theta\theta} - A_{\theta\phi}^2},
\end{equation}
The  $A_{\phi\phi}$, $A_{\theta\theta}$ and $A_{\theta\phi}$ are the
curvatures of the classical energy, defined
as 
\begin{equation}
  E(\vhat{n}) \sim 
\frac{1}{2} A_{\theta\theta} (\theta-\theta_0)^2
+\frac{1}{2} A_{\phi\phi} (\phi-\phi_0)^2
+A_{\theta\phi} (\theta-\theta_0)(\phi-\phi_0) + \cdots,
\end{equation}
where $\phi_0$ is taken to be zero for the $\psi_2$ state.
The cross-term $A_{\theta\phi}$ vanishes, so the classical gap
$\Delta$ is given by the geometric mean of the curvatures divided by
the moment size \cite{jensen-1991-rare,petit-gingras-2014-order}
as in Eq. (\ref{gap-one}). Explicitly,
\begin{equation}
  \label{eq:gap}
  \Delta =  3 \sqrt{C_4^2 + 8 C_2C_6}.
\end{equation}
In this form, we see from Eq. (\ref{eq:endiff}) that the energy
difference, $\delta E$, between the $\psi_3$ and $\psi_2$ states and the gap size are
related as
\begin{equation}
\label{eq:energygap}
\Delta^2 = 36C_2 \delta E.
\end{equation}
We note that in the quantum order by disorder scenario
\cite{savary-balents-2012-order}, a similar relation is
given for the order by quantum disorder gap as found in linear
spin-wave theory, albeit with a different numerical prefactor.  With the
relationship of the multi-spin interactions to the energy difference
and excitation gap in hand, we move on to derive some estimates of the size of 
the four- and six-spin interactions $C_4$ and $C_6$.
\subsection{Scaling arguments}
\label{sec:scaling}
While we have assumed that $|C_4|,|C_6| \ll C_2$, to understand if these
terms are competitive with other effects, such as order by quantum
disorder, we need a more explicit estimate of their magnitude.  First, we
consider simple scaling arguments \cite{savary-balents-2012-order,
petit-gingras-2014-order} for the size of these terms from the strong
coupling expansion sketched in Sec. \ref{sec:strong-coupling}. One has
\begin{align}
  \label{eq:naive-estimates}
  C_2 &\sim \exJ, &
  C_4 &\sim \exJ^3/\cefX{2}, &
  C_6 &\sim \exJ^5/\cefX{4},
\end{align}
where $\mat{\exJ} = \mat{\lambda}^{-1}\mat{\exM} \mat{\lambda}^{-1}$.
The $\mat{\lambda}$-factors of size $\lambda_z \sim 2$ and
$\lambda_{\pm} \sim 6$ provide an accounting for the size of the
matrix elements of the $J=15/2$ operators.  From Eq. (\ref{eq:gap}),
the gap should then scale as $\Delta \sim \exJ^3/\cefX{2}$ when a
$\psi_2$ state has been stabilized.  From the fitted exchanges 
\cite{savary-balents-2012-order,petit-gingras-2014-order}, one has
$\exJ \sim 0.01-0.1 \meV$ and $\cefS \sim 6 \meV$, so even optimistically 
it would seem that one would estimate a very small gap $\Delta \sim 0.001-0.01 \mueV$ via
Eq. (\ref{eq:gap}) and Eq. (\ref{eq:naive-estimates}). This value is
roughly in agreement with that argued in Ref. [\onlinecite{savary-balents-2012-order}],
and suggests that the virtual crystal field excitations are irrelevant to the
ground state selection.

However, this is somewhat na\"ive, as it excludes significant
combinatoric factors.  To see this, consider the multipolar
interactions in $V$ on each nearest-neighbor bond as being independent
perturbations to $\eta H_0$, with $V \equiv \sum_{\avg{ij}}
V_{ij}$. We then define a \emph{contribution} to $\Heff{n}$ as some
string of $n-1$ resolvent operators $R$ and $n$ perturbations
$V_{ij}$. These bonds must be connected and, for the leading
contributions to $C_4$ and $C_6$, must form trees, i.e. have no
loops. Any contribution to $\Heff{n}$ with loops will involve less
than $n$ distinct sites and thus will not be the leading parts of the
coefficient, as mentioned in Sec. \ref{sec:strong-coupling}. 
The number of contributions from three connected
nearest-neighbor bonds of the pyrochlore lattice is of order $10^2$
and from five connected nearest-neighbors is of order
$10^3-10^4$. Since each of these contributions is of order $\exJ$, the
estimates given above should include these large prefactors.  This is
in some sense conservative, as each bond alone has many different
multipolar ranks interacting. We note that the scale $\exM$ is
determined from the fitted constants $\exJ$ and thus is, in some sense,
an overall scale for all the odd-rank multipolar interactions, not just 
those of rank-1.  We thus see that assuming a classical $\psi_2$
selection, when the combinatoric factors are included, the rough
estimate of the excitation gap increases by a factor
of $100$ to $\sim 0.1 - 1\mueV$.

Furthermore, this result is sensitive to the scale of multipolar interactions
between the $J=15/2$ moments; if the scale $\exM$ is increased by some
factor the induced gap increases by its \emph{cube}, as 
shown in Eqs. (\ref{eq:gap}, \ref{eq:naive-estimates}). Moreover, the
above estimate does not include the role played by even-rank
multipolar interactions that do not contribute at first order to the scale
$\exJ$ obtained from the experimental fitting. As we saw in
Eq. (\ref{eq:proj}), these simply give irrelevant constants when
projected into the ground doublets.  Given that there is no reason to
expect the even-rank interactions to be smaller than the odd-rank
\cite{elliott-1968-orbital, santini-2009-multipolar}, this could be a 
\emph{significant underestimate} of the size of the virtual crystal field
fluctuations.  Beyond this, it is not a priori excluded that
the multitude of higher odd-rank
interactions could also have an effect in the virtual crystal field
fluctuations that is disproportionate to their effects in the bare
projection. Since our very rough estimate above is \emph{within an order of magnitude} or so
of the $\sim 20\mueV$ gap estimated from quantum zero-point
fluctuations \cite{savary-balents-2012-order} and the $\sim 40-45 \mueV$
gap measured experimentally \cite{ross-gaulin-2014-order,
petit-gingras-2014-order}, the importance of virtual crystal
fluctuations must be re-evaluated. These heuristic arguments are one of the
main results of this work, forming the backbone of the detailed calculations we
present next. In sections to follow, we flesh out these statements with 
quantitative calculations of the energetic selection and excitation gap
due to the effects of virtual crystal field fluctuations.

\section{Bilinear-biquadratic model}
\label{sec:biquadratic} 
As argued in the previous section, simple scaling arguments suggest that
the effects of virtual crystal field fluctuations may be significant
in \eto{}.  To address the magnitude of these effects in a
quantitative fashion, we first need an explicit model of the
multipolar interactions within the $J=15/2$ manifold, rather than the
schematic form of the previous section.  Determination of these
multipolar interactions from experiments is a difficult endeavor; as
most of the experimental data on \eto{} can be described almost
quantitatively using the effective spin-1/2 model of
Ref. [\onlinecite{savary-balents-2012-order}], there is little data
left to constrain the model of Eq. (\ref{eq:j-model}).  Indeed, the
odd-rank interactions are only constrained through their projection
into the effective spin-1/2 model of Eq. (\ref{eq:s-model}), while no
constraints at all are placed on the even-rank interactions.
Estimating the multipolar couplings by purely theoretical means is
also similarly challenging. Given this daunting state of affairs, we
will not attempt to study the full $J=15/2$ multipolar interactions,
but will restrict to some simpler caricature that captures the
essential physics of having both time-reversal even and odd multipolar
interactions. 

The most straightforward approach
\cite{petit-gingras-2014-order,mcclarty-2009-energetic} is to consider
only bilinear interactions between the $J=15/2$ moments. This takes
the form
\begin{equation}
  \label{eq:bilinear-model}
H_{J} \equiv  \sum_{\avg{ij}} \trp{\vec{\mJ}}_i\left[
    \mat{\lambda}^{-1} \mat{\exJ_{ij}} \mat{\lambda}^{-1}
\right] \vec{\mJ}_j +
      \eta \sum_i \cfV(\vec{\mJ}_i).
\end{equation}
The exchanges $\mat{\lambda}^{-1} \mat{\exJ}_{ij} \mat{\lambda}^{-1}$
have been chosen to reproduce Eq. (\ref{eq:s-model}) after one
projects into the ground doublet, with $P(\mat{\lambda}^{-1}
\vec{\mJ}_i)P = \vec{\mS}_i$.  While appealing in its simplicity, the
lack of even-rank interactions or higher odd-rank interactions
eliminates a possibly potent source of virtual crystal field
fluctuations. To capture some aspects of these even-rank multipolar
interactions that must be present in the microscopic model, we
supplement this bilinear model with \emph{biquadratic} interactions
between $J=15/2$ moments. These are couplings between rank-2
multipoles as defined in Eq. (\ref{multipole}) with
\begin{equation}
  H_Q \equiv \sum_{\avg{ij}} \sum_{QQ'}  \exM^{2Q,2Q'}_{ij} O_{2,Q}(\vec{\mJ}_i) O_{2,Q'}(\vec{\mJ}_j),
\end{equation}
where $Q,Q' = 0,\pm 1,\pm 2$. The interaction matrix needs
only be specified on a single bond
while the rest can be determined by symmetry.
While in principle unconstrained, we expect the scale of these interactions to be
comparable to that of the bilinear terms. 
We thus consider the \emph{bilinear-biquadratic} model
\begin{align}
  \label{eq:jq-model}
  H &\equiv H_J + \kappa H_Q + \eta H_{\rm CEF}, \nonumber \\
&= \sum_{\avg{ij}} \trp{\vec{\mJ}}_i\left[
    \mat{\lambda}^{-1} \mat{\exJ_{ij}} \mat{\lambda}^{-1}
\right] \vec{\mJ}_j +
      \eta \sum_i \cfV(\vec{\mJ}_i) \nonumber \\
&+
\kappa \sum_{\avg{ij}} \sum_{QQ'}  \exM^{2Q,2Q'}_{ij} O_{2,Q}(\vec{\mJ}_i) O_{2,Q'}(\vec{\mJ}_j).
\end{align}
We have added a single tuning parameter $\kappa$, in analogy to the parameter
$\eta$, to control the scale of the biquadratic interactions. To ensure
we are treating the rank-1 and rank-2 interactions on equal footing, we
normalize the matrix $\mat{\exM}^{2,2}_{ij}$ so that 
\begin{equation}
  \label{eq:norm}
  \frac{1}{5^2} \sum_{QQ'} |\exM^{2Q,2Q'}_{ij}|  = \frac{1}{3^2} \sum_{QQ'} |\exM^{1Q,1Q'}_{ij}|,
\end{equation}
where $\mat{\exM}^{11}_{ij}$ are the spherical tensor components of the rank-1 interactions,
as normalized in Eq. (\ref{multipole}). In this way, the average matrix elements for
the rank-1 and rank-2 interactions are equal when $\kappa=1$.

In this model, the bilinear terms serve as a proxy for \emph{all} the odd-rank
interactions, while the biquadratic terms do the same for \emph{all} the even-rank
interactions. Even with this simplification, there are a large number of
free parameters.  For nearest-neighbor bonds we find that there are 10 symmetry
allowed independent biquadratic couplings. We will fix the ratios of
these couplings using models motivated from two microscopic schemes,
leaving the overall scale $\kappa$ as a tuning parameter. These schemes are
complementary and motivated by super-exchange \cite{onoda-2011-quantum-fluctuations} and
the microscopic electric quadrupole-quadrupole interaction
\cite{wolf-birgeneau-1968-electric}.  We do \emph{not} intend to imply that
these different schemes are strictly realistic models for biquadratic exchange
in \eto{}, though both would be present in the material;
 they simply serve as useful parametrizations of the
biquadratic couplings. Furthermore, the similarity of the results for the
different biquadratic coupling schemes that we find below is a strong indication that the
order by virtual crystal field fluctuation mechanism is enhanced when including
biquadratic couplings, \emph{independently} of the details of even-rank
interactions.

\subsection{Super-exchange model}
\label{sec:super-exchange}
One natural source of higher-multipole interactions \footnote{
Note that due to the large orbital moments in the Er\tsup{3+} ground
doublets, super-exchange generates multipolar interactions at leading
order. This is in contrast to spin-only moments where they appear only
at higher order in perturbation theory.
} in rare-earths
comes from super-exchange \cite{santini-2009-multipolar}.  Due to the
large separation of the Er\tsup{3+} ion relative to the extent of the
$4f$ orbitals, we expect this proceeds largely through the neighboring
oxygen atoms. For a detailed treatment we refer the reader to
Refs. [\onlinecite{onoda-2011-quantum-fluctuations,rau-2015-quantum-ice}]. 
Here, we simply quote the form of
the interaction in the so-called charging approximation \footnote{
In this approximation, only the energy differences between atomic 
states with different charge configurations are retained. The splittings of the
atomic manifold within each $4f^n$ configuration are ignored.
}:
\begin{equation}
\sum_{\avg{ij}}
 \sum_{\alpha\beta\mu\nu}
\left(P \h{f}_{i\alpha} f^{}_{i\beta} P\right) 
\mathcal{I}^{\alpha\beta\mu\nu}_{ij}
\left(P \h{f}_{j\mu} f^{}_{j\nu}  P\right),
\end{equation}
where we have used a combined spin-orbital
index $\alpha \equiv (m,\sigma)$ and
$\h{f}_{i\alpha}$ creates a $4f$ electron with spin $\sigma$ and orbital $m$
at site $\vec{r}_i$.
The local operators can be expressed in terms of multipoles of rank-$K$
with $K \leq 7$
\begin{equation}
P\h{f}_{i\alpha} f^{}_{i\beta} P \equiv
 \sum_{KQ} A_{KQ}^{\alpha\beta} O_{KQ}(\vec{\mJ}_i),
\end{equation}
for some set of coefficients $A_{KQ}^{\alpha\beta}$.  The interaction
matrix $\mathcal{I}$ is defined as
\begin{eqnarray}
\label{eq:super-exchange-model}
&&\mathcal{I}^{\alpha\beta\mu\nu}_{ij} \equiv
\frac{2}{(U^+_f +\Delta_{fp})^2} 
\Bigg[
- \frac{2
\left[
{t}^{}_{j}
\h{t}_{j}
\right]^{\mu\nu}
\left[
{t}^{}_{i}
\h{t}_{i}\right]^{\alpha\beta}
}{2U^+_f +U_p +2\Delta_{fp}} +
\nonumber \\
&& 
\left(\frac{1}{U^+_f+U^-_f}+\frac{2}{2U^+_f +U_p +2\Delta_{fp}}\right)
\left[{t}^{}_{j} \h{t}_{i} \right]^{\mu\beta}
\left[{t}^{}_{i} \h{t}_{j} \right]^{\alpha \nu}
\Bigg],
\end{eqnarray}
where $U^{\pm}_f \sim E(4f^{11 \pm 1}) -  E(4f^{11})$ gives the energy
differences between adjacent charge configurations of the Er\tsup{3+} ion.
$\Delta_{fp}$ is the charge gap between the $4f$ and $2p$ levels.  The
$t_i$ matrices define the $4f$-$2p$ overlap. In the Slater-Koster
approximation \cite{slater-koster-1954-simplified-lcao}, these are
expressed as $ t_{i} = \h{R}_{i} t_0 $ where $R_{i}$ is a rotation of
the $f$ and $p$ orbitals that takes a set of axes aligned to the local
axes at site $\vec{r}_i$ into the global frame.  The matrices $t_0$
define overlaps in the frame aligned along the bond axis and take the
simple form $[t_0]_{mm'} = \delta_{mm'} (\delta_{|m|=1} t_{pf\pi} +
\delta_{m=0} t_{pf\sigma})$.

\begin{figure}[tp] 
  \centering
  \includegraphics[width=\columnwidth]{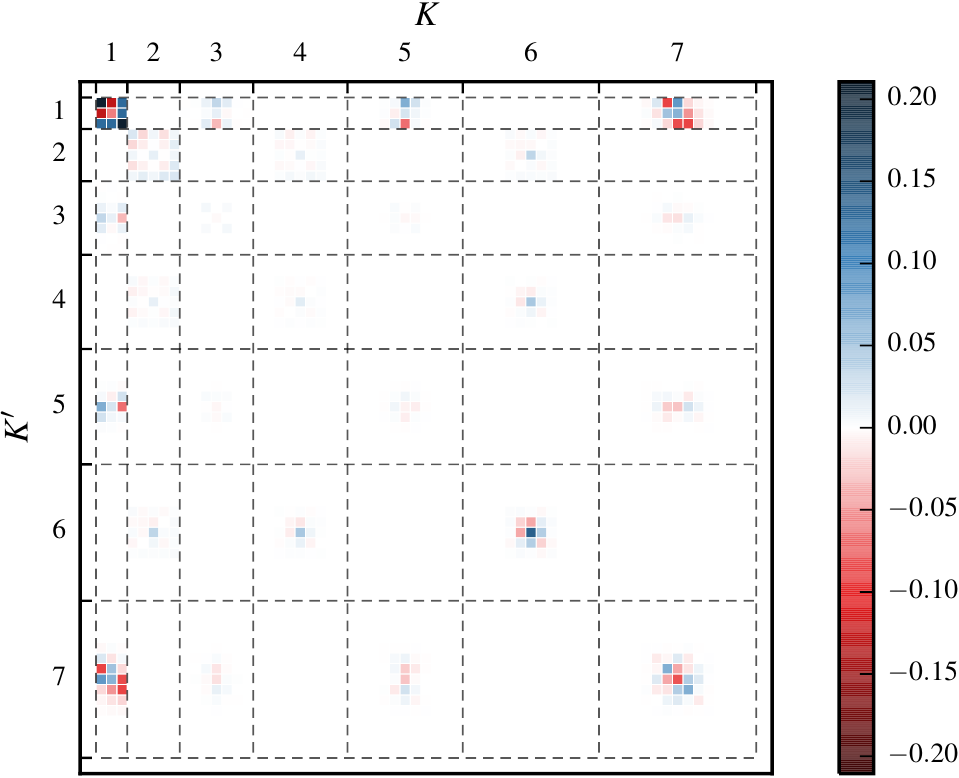}
  \caption{\label{fig:multipolar-int}
Representation of the multipolar interaction matrix $\mat{\exM}$
generated by the super-exchange interaction given in
Eq. (\ref{eq:super-exchange-model}). 
The overall scale is arbitrary.
For each pair of ranks $K,K' =
1,2,\ldots, 7$ the matrix elements $\exM^{KQK'Q'}$ can be arranged into
a $2K'+1$ by $2K+1$ matrix with the $Q,Q'$ indices running along the
column and row respectively. These blocks are shown arranged in a
table by their $K,K'$ indices, with the dashed lines showing the
boundaries of each block where $Q = \pm K$ or $Q' = \pm K'$.
Significant interactions beyond purely bilinear, i.e. with $K, K' > 1$, are present.
  }
\end{figure}

Since $U_p >0$ and $\Delta_{fp}>0$, the second term in
Eq. (\ref{eq:super-exchange-model}), with denominator
$(U_f^++U_f^-)^{-1}$, is dominant.
Taking only this term
affords the simplification that the parameters $U^{\pm}_f$ and
$\Delta_{pf}$ enter only into the overall scale of the super-exchange
interactions.  Fixing the Slater-Koster parameters to the canonical
ratio \cite{takegahara-1980-slater-koster} $t_{pf\pi}/t_{pf\sigma}
\sim -0.3$, one finds that a very complex set of multipolar
interactions are generated. As illustrated in
Fig. \ref{fig:multipolar-int}, these include significant odd and
even-rank interactions up to and including rank-7
\cite{iwahara-2015-exchange, rau-2015-quantum-ice}. To fit this within
our bilinear-biquadratic framework, we simply truncate all multipolar
interactions with rank $> 2$ and replace the rank-1 interactions with
those of Eq. (\ref{eq:jq-model}) to reproduce the experimentally
fitted model of Eq. (\ref{eq:s-model}) when projected into the ground
doublet. We again emphasize that this is not expected to be a
quantitative model of the microscopic super-exchange interactions due
to the neglect of the higher multipoles as well as the charging
approximation used to arrive at
Eq. (\ref{eq:super-exchange-model}). However, we \emph{do} expect this
form to capture the essential features of even-rank interactions and
thus will serve as a useful tool for exploring how these multipolar
interactions affect the virtual crystal field fluctuations.

\subsection{Electric quadrupole-quadrupole model}
\label{sec:eqq}
Another simple interaction between the quadrupolar degrees of freedom
arises simply due to electrostatics \cite{finkelstein-mencher-1953,
bleaney-1961,wolf-birgeneau-1968-electric}.  To present this
compactly, it will be useful to consider a more conventional
definition of the quadrupolar operators, rather than the spherical
tensor basis we have been using so far.  We consider the Cartesian
tensors $\vec{Q}_{ij}$ defined with respect to the \emph{global} axes
as
\begin{equation}
\label{quadrupoles}
  \vec{Q}_i \equiv \trp{\vec{R}}_i\left( 3 \overline{\vec{J}^{}_i \trp{\vec{J}}_i} - J(J+1)\right) \vec{R}_i,
\end{equation}
where $\trp{\vec{R}}_i {\vec{J}}^{}_i \equiv \vhat{x}^{}_i J^x_i +
\vhat{y}^{}_i J^y_i + \vhat{z}^{}_i J^z_i$ is the moment at site
$\vec{r}_i$ in the global axes and the product is symmetrized,
i.e. $\overline{J^{\mu} J^{\nu}} \equiv (J^{\mu} J^{\nu}+ J^{\nu} J^{\mu})/2$.  This
is a symmetric traceless matrix with five independent components. The
explicit relationship between spherical tensors defined in
Eq. (\ref{multipole}) and the Cartesian form of
Eq. (\ref{quadrupoles}) is given Appendix \ref{cartesian}.  These
electric quadrupole-quadrupole (EQQ) interactions decay as $\sim
1/r^5$ so we consider only the nearest-neighbor contribution
\begin{equation}
\mathscr{Q}  \sum_{\avg{ij}} 
  \trp{\vhat{r}}_{ij}\left(
    \frac{1}{3} \tr\left[\vec{Q}_i \vec{Q}_j\right]-
    \frac{10}{3} \vec{Q}_i \vec{Q}_j
    +\frac{35}{6}
    \vec{Q}_i \vhat{r}_{ij}
    \trp{\vhat{r}}_{ij}  \vec{Q}_j
  \right)\vhat{r}_{ij},
\end{equation}
where $\vec{r}_{ij} \equiv \vec{r}_i - \vec{r}_j$. The microscopic quadrupolar
coupling constant $\mathscr{Q}$ is given by
\cite{wolf-birgeneau-1968-electric}
\begin{equation}
  \mathscr{Q} \equiv \frac{\alpha_J \avg{r^2}^2}{r^5_{\rm nn}} \left(\frac{e^2}{4\pi\epsilon}\right)
  \sim 6.664 \times 10^{-6} \meV.
\end{equation} 
Shielding effects likely reduce this further by a factor of $\sim
0.25$ or so \cite{wolf-birgeneau-1968-electric}. The smallness of this
numerical value is highly misleading; given the large matrix elements
$\sim J^4$ the true scale of these interactions is closer to $10^{-3}
\meV$ to $10^{-4} \meV$. As for the super-exchange model, this
interaction is used as a convenient parametric form of the biquadratic couplings and
is not intended to be realistic.  We thus do not use the bare
microscopic interaction strength, but normalize the biquadratic
couplings as given in Eq. (\ref{eq:norm}), allowing the overall scale
to vary via $\kappa$.

\subsection{Mean-field and random-phase approximation}
\label{sec:rpa}
With these concrete models defined, we can
now analyze the effects of the biquadratic interactions within an approach
that uses mean-field theory (MFT) and the random-phase approximation (RPA)
 \cite{jensen-1991-rare}.  
For the case of purely
bilinear interactions, such an analysis has been presented in
Ref. [\onlinecite{petit-gingras-2014-order}] (differences from the
present study are discussed in Appendix \ref{app:diff}).  This provides a route
to estimating the gap, and thus the condensation energy, without the
heavy machinery of the high-order perturbative expansion presented in
Sec. \ref{sec:strong-coupling}.  We lose however, the quantitative mapping to
the effective spin-1/2 degrees of freedom. Still, as we shall see, the
MFT and RPA results are entirely consistent with the scaling expected from
the strong coupling expansion, and even include effects that go beyond the
leading order results given in Sec. \ref{sec:strong-coupling}.
We formulate these
methods for general multipolar interactions, then apply the results to
the specific bilinear-biquadratic models discussed in
Secs. \ref{sec:super-exchange} and \ref{sec:eqq}

In the mean-field approximation we simply decouple the multipolar
interactions as
\begin{align}
  O_{KQ}(\vec{\mJ}_i)  O_{K'Q'}(\vec{\mJ}_j) &\approx
\mathscr{O}_{j}^{K'Q'} O_{KQ}(\vec{\mJ}_i)+ 
 \mathscr{O}^{KQ}_i  O_{K'Q'}(\vec{\mJ}_j) \nonumber\\
& -  \mathscr{O}^{KQ}_i \mathscr{O}^{K'Q'}_j,
\end{align}
where $\mathscr{O}^{KQ}_i \equiv \avg{O_{KQ}(\vec{\mJ}_i)}$ are
expectation values of the multipole operators. The full multipolar
exchange Hamiltonian then reduces to single-site terms
\begin{align}
  \label{eq:single-site}
  H \approx \sum_i \left[
  \eta \cfV(\vec{\mJ}_i)
  + \sum_{KQ} h_{KQ,i} O_{KQ}(\vec{\mJ}_i)
  \right] \equiv \sum_i \mathscr{H}_i(\vec{\mJ}_i),
\end{align}
where we have defined the effective multipolar mean fields
\begin{equation}
h_{KQ,i} \equiv \sum_j\sum_{K'Q'} \left(
\exM^{KQ,K'Q'}_{ij} +\exM^{K'KQ'Q}_{ji}\right)
 \mathscr{O}^{K'Q'}_j.
\end{equation}
This defines a set of effective single-site Hamiltonians
$\mathscr{H}_i$ that can be solved self-consistently for the multipole
expectation values $\mathscr{O}^{KQ}_i$. This is equivalent
to performing variational minimization on an ansatz that is an
arbitrary product state in the $J=15/2$ site basis.

To access the spin-wave excitation gap, $\Delta$, we compute the dynamic
susceptibility $\mat{\chi}(\vec{k},\omega)$ within the RPA.  The
imaginary part of $\mat{\chi}(\vec{k},\omega)$ is directly related to
the intensity observed in inelastic neutron scattering experiments.
With only bilinear interactions between the 
$\vec{J}_i$ moments, one can use the standard 
RPA equation \cite{jensen-1991-rare}
\begin{equation}
  \left[1-\mat{\chi}^0(\omega) \mat{\exJ}(\vec{k})\right] \mat{\chi}(\vec{k},\omega) = 
  \mat{\chi}^0(\omega),
\end{equation}
where $\mat{\exJ}(\vec{k})$ is the Fourier transform of the bilinear exchange matrix
and $\mat{\chi}^0(\omega)$ is the on-site susceptibility
\begin{equation*}
\chi^0_{\mu\nu} (\omega) \equiv
  \sum_{nn'}
    \frac{\bra{n}{\mJ_{\mu}}\ket{n'}
      \bra{n'}{\mJ_{\nu}}\ket{n}}
    {(\omega + i0^+)-(E_{n'}- E_n)}
    \left(\rho_{n'} - \rho_{n}\right),
\end{equation*}
where $E_n$ and $\ket{n}$ denote the energies and eigenstates the
single-site mean-field Hamiltonian defined in Eq. (\ref{eq:single-site}) and
$\rho_n \equiv e^{-\beta E_n}/Z$ are the associated Boltzmann weights.
 
With multipolar interactions, we compute instead a generalized
multipolar susceptibility $\mat{\Gamma}(\vec{k},\omega)$.
As the derivation follows that of the conventional RPA
closely \cite{jensen-1991-rare}, we simply state the final expressions.
We first define the on-site, non-interacting multipolar susceptibility
\begin{equation}
  \Gamma^0_{KQ,K'Q'} (\omega) \equiv
  \sum_{nn'}
    \frac{\bra{n}{O_{KQ}(\vec{\mJ})}\ket{n'}
      \bra{n'}{O_{K'Q'}(\vec{\mJ})}\ket{n}}
    {(\omega + i0^+)-(E_{n'}- E_n)}
    \left(\rho_{n'} - \rho_{n}\right),
\end{equation}
This is can be regarded as a matrix $\mat{\Gamma}^0(\omega)$
in the basis of spherical tensors, with indices $(KQ)$ and
$(K'Q')$. The magnetic susceptibility $\mat{\chi}(\vec{k},\omega)$ is
simply the $\Gamma_{1Q,1Q'}(\vec{k},\omega)$ block of
$\mat{\Gamma}(\vec{k},\omega)$ rotated into the Cartesian basis using
Eq. (\ref{eq:vectortensor}). In the RPA, the full wave-vector
dependent multipolar susceptibility $\mat{\Gamma}(\vec{k},\omega)$
then satisfies the equation
\begin{equation}
  \left[1-\mat{\Gamma}^0(\omega)\mat{\exM}(\vec{k})\right] 
  \mat{\Gamma}(\vec{k},\omega) =  \mat{\Gamma}^0(\omega),
\end{equation}
where $\mat{\exM}(\vec{k})$ is the Fourier transform of the multipolar
interaction matrix $\mat{\exM}_{ij}$. As both $\mat{\exM}(\vec{k})$
and $\mat{\Gamma}^0(\omega)$ are known, these are linear equations
that can be solved for each pair $(\vec{k},\omega)$ to yield
$\mat{\Gamma}(\vec{k},\omega)$. To cure the singularities, a small
imaginary part is added to denominators in $\mat{\Gamma}^0(\omega)$,
endowing the excitation spectrum with a slight broadening.
For the bilinear-biquadratic model, Eq. (\ref{eq:jq-model}), that we
are studying, it is sufficient to consider
$\mat{\Gamma}(\vec{k},\omega)$ as an eight by eight matrix that
includes $\mat{\chi}(\vec{k},\omega)$, a five by five quadrupolar
susceptibility, as well as off-diagonal bilinear-biquadratic mixing
terms.

\section{Results}
\label{sec:results}
\begin{figure}[tp]
  \centering
  \includegraphics[width=0.6\columnwidth]{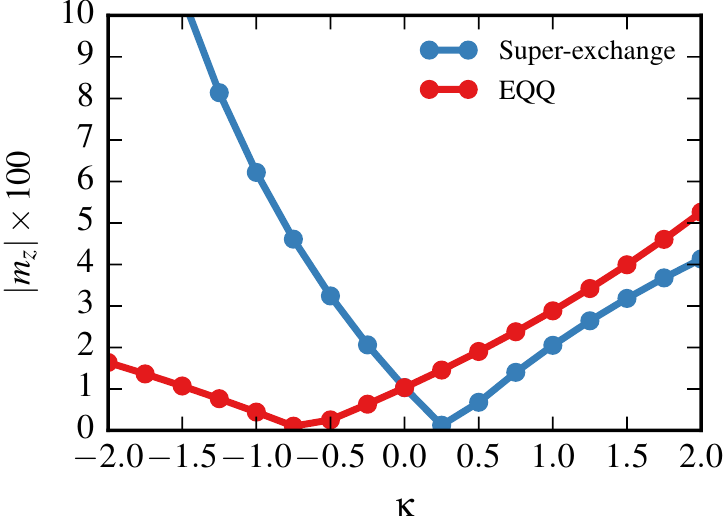}
  \includegraphics[width=0.33\columnwidth]{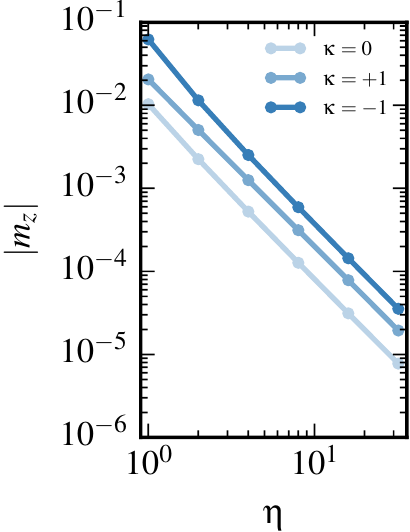}
  \caption{\label{fig:mz} (a) Dependence of the canted moment $m_z$
    on the biquadratic coupling strength $\kappa$ in 
    the super-exchange and EQQ cases. Note that
    throughout the range $|m_z|$ is at most $\sim 0.1$ while the 
    in-plane components $m_{\perp}$ take on a value $\sim 3$ expected
    from the effective-spin picture. (b) Dependence of
    $|m_z|$ on the crystal field rescaling $\eta$, showing the $\eta^{-2}$
    scaling for several values of $\kappa$ for the super-exchange model.
  }
\end{figure}

The MFT is formulated in terms of the
order parameters for rank-1 and rank-2 multipoles, defined as
\begin{align}
  \vec{m}_i &\equiv \avg{\vec{\mJ}_i}, &
  q_{Q,i} &\equiv \avg{O_{2,Q}(\vec{\mJ}_i)}.
\end{align}
The five expectation values $q_{Q,i}$ can be organized as a vector
$\vec{q}_i \equiv (q_{+2,i}\ q_{+1,i}\ q_{0,i}\ q_{-1,i}\ q_{-2,i})$.
We note that the $q_{0,i}$ component is non-zero even without any
ordering as $O_{2,0}(\vec{\mJ}_i)$ breaks no symmetries of the model.
Concretely, this is due to the presence of the crystalline electric field.
Solving the self-consistent equations on a single tetrahedron
for both $\vec{m}$ and $\vec{q}$, we generically find a $\psi_2$ state
with uniform mean fields $\vec{m}_i \equiv \vec{m}$ and $\vec{q}_i \equiv \vec{q}$. 
In contrast to the effective spin-1/2 model of
Eq. (\ref{eq:s-model}), the degeneracy of the $\Gamma_5$ manifold is
already lifted at the mean-field level through the inclusion of the
higher crystal field levels, as previously found in 
Refs. [\onlinecite{mcclarty-2009-energetic}, \onlinecite{petit-gingras-2014-order}].
In this local basis, the $\psi_2$ states are characterized by the
expectation value $\vec{m}$, taking one
of the six symmetry related solutions
\begin{subequations}
\label{eq:domains}
\begin{eqnarray}
  \vec{m} 
  &=& \pm m_{\perp} \vhat{x} \mp m_z \vhat{z}, \\
  &=& 
\pm m_{\perp}\left(-\frac{1}{2}\vhat{x} \pm \frac{\sqrt{3}}{2} \vhat{y}\right) 
\mp m_z \vhat{z}, \\
  &=& 
\pm m_{\perp}\left(-\frac{1}{2}\vhat{x} \mp \frac{\sqrt{3}}{2} \vhat{y}\right) 
\mp m_z \vhat{z},
\end{eqnarray}
\end{subequations}
with $m_{\perp} \sim O(1)$ and $|m_z| \ll m_{\perp}$ at low temperature.
One can think of these as planar $\psi_2$ states with a small amount
of canting into the local $\vhat{z}$ direction, as expected from the
analysis of Sec. \ref{sec:classical}. 
The small component $m_z$ is a secondary order parameter
and is pinned as $m_z \propto m^3_{\perp}$ as temperature is
varied. This can be understood at the level of Landau-Ginzburg theory
where allowed terms such $\sim m_z m_{\perp}^3$ cause $m_{\perp}^3$ to
act as a conjugate field to $m_z$ in the ordered phase
\cite{javanparast-gingras-2015-order}.  As $m_z$ is induced by the
four-spin term $C_4$, it gives some indication of the magnitudes of
energetic selection and excitation gap. However, as we saw in Sec. \ref{sec:classical},
both these effects depend on not only $C_4$ but on the six-spin term $C_6$ as well.
Since the small $m_z \sim \theta_0$ 
moment is \emph{independent} of $C_6$ (see Eq. (\ref{eq:mz})),
the magnitude of $|m_z|$ gives only a partial picture of the selection mechanism.
This is particularly true in regions where $m_z$, and thus $C_4$, change sign
relative to $m_{\perp}$.

 The exact values of $m_{\perp}$ and $m_z$ depend on the crystal field
and exchange parameter sets used as well as the strength of the
biquadratic coupling. For definiteness, we focus on the crystal
field parametrization of \citet{bertin-2012-crystal} with the bilinear
exchange parameters of \citet{savary-balents-2012-order} (see
Appendix \ref{parameters}) mapped to bilinear interactions
as in Eq. (\ref{eq:bilinear-model}).
 Results using other parameter sets (see Appendix \ref{parameters} ) from
the literature give qualitatively similar values. We then can gain some insight into the effects
of the biquadratic interactions by looking at the dependence of $|m_z|$
on the tuning parameter $\kappa$ at $T=0$.  We present these results for the
super-exchange and EQQ model interactions in Fig. \ref{fig:mz}.  For
$\kappa=0$ one find that $|m_z|$ is small but finite, of order $\sim
10^{-2}$.  This is purely a consequence of the high-lying crystal
field states, as can be seen by looking at the dependence of $|m_z|$ on
the crystal field rescaling $\eta$. As shown in Fig. \ref{fig:mz}, it
scales as $\sim \eta^{-2}$ as expected from Sec. \ref{sec:classical}.
Turning to the effect of finite $\kappa$, we see that the magnitude of
$m_z$ increases dramatically from the bare value $|m_z| \sim 10^{-2}$
present at $\kappa = 0$. This indicates generation of four-spin term
and thus that the selection of the $\psi_2$ state can be significantly
enhanced by including biquadratic interactions, or more generally,
high-rank multipolar interactions.

\begin{figure}[tp]
  \centering
  \includegraphics[width=0.6\columnwidth]{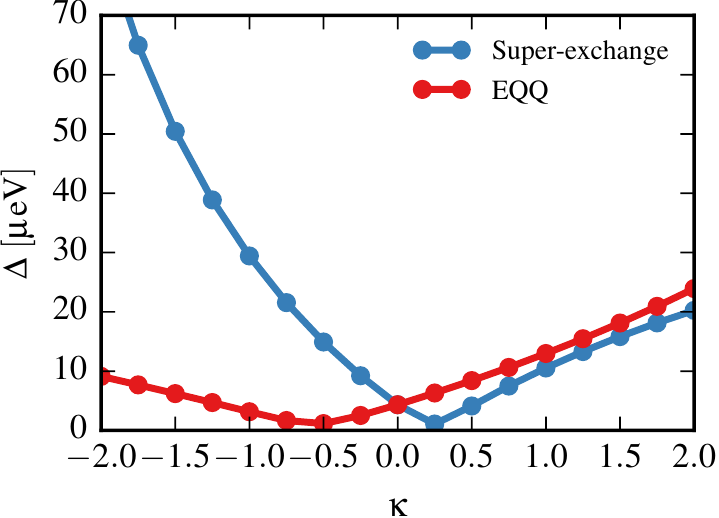}
  \includegraphics[width=0.33\columnwidth]{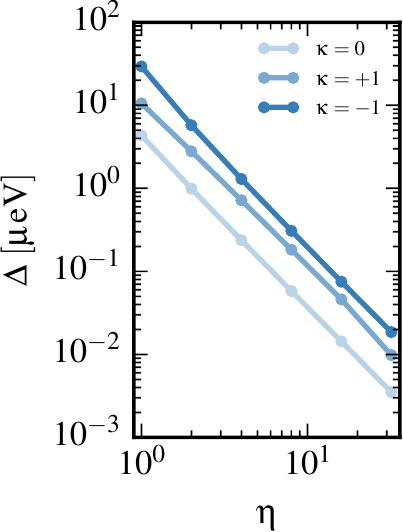}
  \caption{\label{fig:gap} (a) Dependence of the gap excitation gap
    $\Delta$ on the biquadratic coupling strength $\kappa$ for
    super-exchange and EQQ type interactions.  Note that for $\kappa \sim
    2$ both super-exchange and EQQ give a gap of $\sim 20 \mueV$,
    comparable to that found for order by quantum disorder.  For the
    super-exchange model with $\kappa \sim -1.5$, the gap approaches the
    experimental value of $\sim 40-45 \mueV$.
    (b) Dependence of $\Delta$ on the crystal field rescaling $\eta$,
    showing the $\eta^{-2}$ scaling for several values of $\kappa$ for the
    super-exchange model.
  }
\end{figure}

\begin{figure}[tp]
  \centering
  \includegraphics[width=0.9\columnwidth]{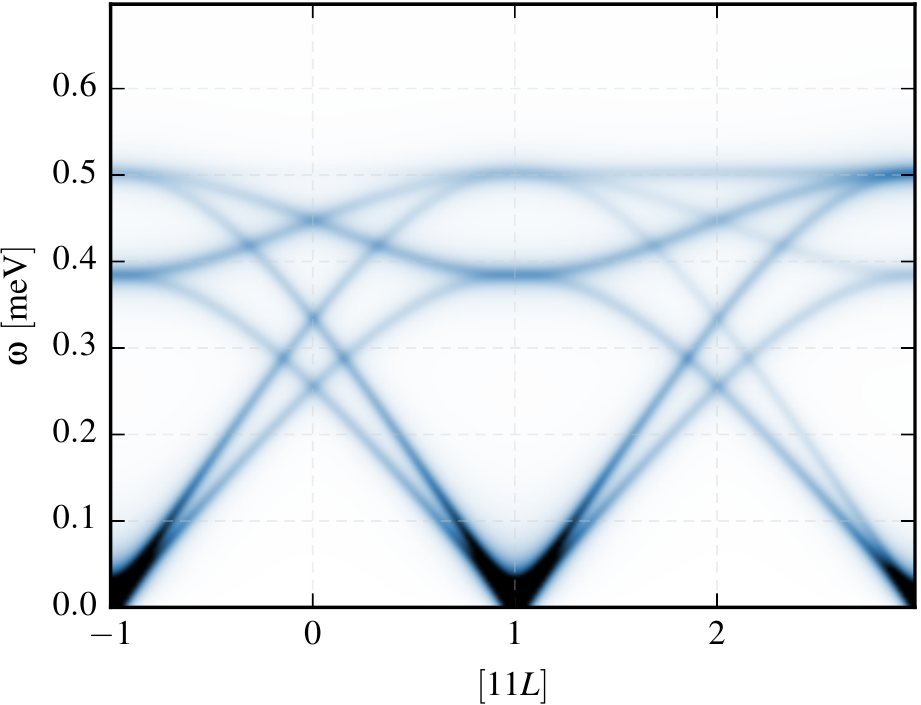}
  \caption{\label{fig:spectrum-no-gap}
    Excitation spectrum of the effective spin-1/2 model ($\eta \rightarrow \infty$) in the RPA,
    cut along the $[11L]$ direction. Intensity scale is arbitrary. The gapless
    mode is visible near $[111]$.
  }
\end{figure}

\begin{figure}[tp]
  \centering
  \includegraphics[width=0.9\columnwidth]{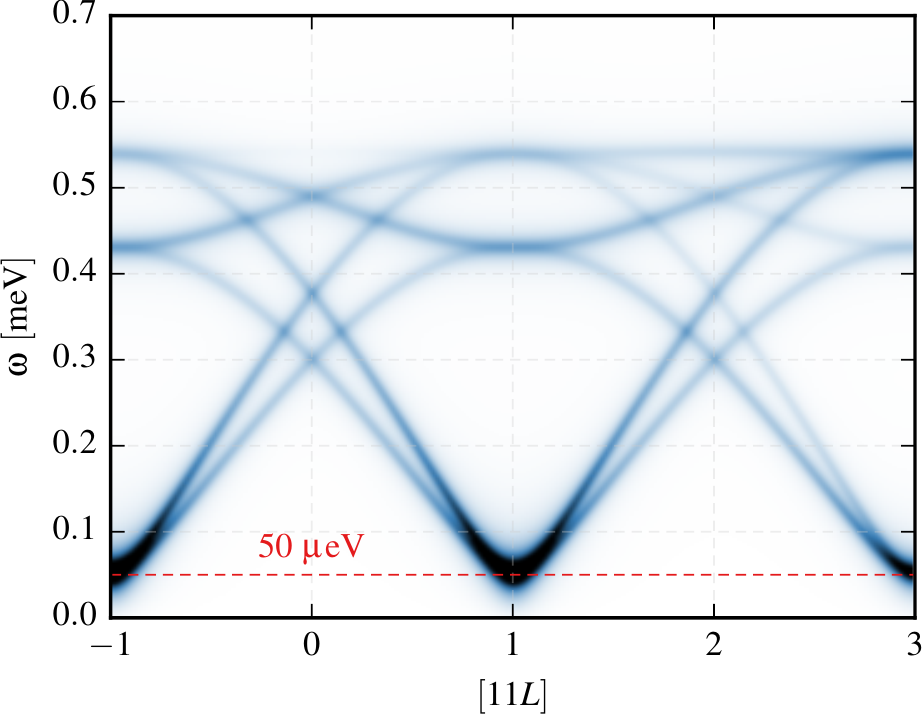}
  \caption{\label{fig:spectrum-gap}
    Excitation spectrum of the bilinear-biquadratic model ($\eta = 1$) in the RPA,
shown for super-exchange interactions with $\kappa=-1.5$ along the
$[11L]$ direction. Intensity scale is arbitrary.  
Aside from the $\sim 50 \mueV$ gap near $[111]$, the bands show
is a very strong similarity to the spin-1/2 case ($\eta \rightarrow \infty$) in
Fig. \ref{fig:spectrum-no-gap}.
  }
\end{figure}

Experimentally, while the selection itself can be observed \cite{poole-wills-2007,champion-harris-2003}, there is no
good way to probe the condensation energy, $\delta E$, directly.  Instead, we look
to the gap in the excitation spectrum via the RPA.  As to more closely
compare with the experimental results, we present dynamic structure
factor as would be seen in inelastic neutron scattering experiments
\cite{jensen-1991-rare}
\begin{equation}
  S(\vec{k},\omega) \propto \sum_{\mu\nu}
  \left(\delta_{\mu\nu} - \frac{k^\mu k^\nu}{|\vec{k}|^2}\right)
  \frac{\im{\chi^{\mu\nu}(\vec{k},\omega)}}{1-e^{-\beta\omega}},
\end{equation}
which is directly related to the results of inelastic neutron
scattering experiments \cite{
savary-balents-2012-order,zhitomirsky-moessner-2012-order}. 
We average over the six domains of the
$\psi_2$ ordering, as given in Eq. (\ref{eq:domains}), to compare
directly with experimental results. For simplicity, we do not include
the magnetic form factor of Er\tsup{3+}. The gap size as a function of
$\kappa$ is shown in Fig. \ref{fig:gap}. As in the case of $m_z$, the
scaling of the gap with the crystal field is also consistent with the
order by virtual-crystal field fluctuations scenario; the gap closes
as $\eta^{-2}$, as expected from Sec. \ref{sec:classical}.  The minima
as a function of $\kappa$ are those points where $m_z$, and thus
$C_4$, change sign relative to $m_{\perp}$, and selection is purely from $C_6$. As shown
in Fig. \ref{fig:gap}, the excitation gap is \emph{strongly dependent} on the
strength of the biquadratic coupling; for moderate, but reasonable values of $\kappa$,
one can find a gap that is comparable \emph{or larger} than that found
in the order by quantum-disorder scenario \cite{savary-balents-2012-order}.  
In fact, when the
biquadratic coupling is comparable to the bilinear, this gap can easily
be made consistent with experimental estimates. For the
super-exchange model, this occurs near $\kappa \sim -1.5$, while for
the EQQ model this is at the larger value $\kappa \sim 3.5$.  Note
that the gross features of the excitation spectrum are preserved when
a finite biquadratic coupling is introduced. For concreteness, we compare
the $\eta \rightarrow \infty$ limit where our model reduces to an
effective spin-1/2 model as shown in Fig. \ref{fig:spectrum-no-gap} to
the result with finite super-exchange type biquadratic interactions
shown in Fig. \ref{fig:spectrum-gap}. The biquadratic coupling has
been chosen to be $\kappa = -1.5$ to reproduce the experimental gap
size. Aside from the gap at $[111]$ and small upward shift
in energy, the features of the two spectra
are nearly indistinguishable. Comparing to the experimental results of
Ref. [\onlinecite{savary-balents-2012-order}] or Ref. [\onlinecite{petit-gingras-2014-order}] we find that, aside from
the gap, the model with the biquadratic coupling fits the data as well
as the effective spin-1/2 model within theoretical and experimental
uncertainties.  This agreement should be further improved by
renormalizing the bare $\mat{\exJ}$-couplings to remove the
second-order shifts from virtual crystal field fluctuations.  This
qualitative similarity persists past the point $\kappa=1$, where we
consider the bilinear and biquadratic interactions to be of comparable
magnitude.  Indeed, we can extend the EQQ model to the larger value
$\kappa \sim 3.5$ and still obtain a spectrum nearly identical to
that shown in Fig. \ref{fig:spectrum-gap}.  This indicates that there
is a wide range of biquadratic coupling strengths that induce
a significant gap and remain consistent
with the overall features of the excitation spectrum as reported
in Ref. [\onlinecite{savary-balents-2012-order}].

These two consequences of the biquadratic interactions, the large
induced gap and the invariance of the broad features of the spectrum
form the main result of this work. Together, they imply that
significant biquadratic coupling can not only produce a gap large
enough to account for the observed results but, crucially, can do so without
spoiling the agreement between the excitation spectrum and
experiments. More generally, this tells us that the higher multipolar
interactions, that are expected to be not only present but
significant \cite{santini-2009-multipolar}, 
can have observable effects at low energies. With this
proof of principle, we see that virtual crystal field effects can not necessarily 
be ignored. Next we discuss how this scenario could be tested
in \eto{}.

\section{Discussion}
\label{sec:discussion}
The question of the role of order by disorder type mechanism in \eto{}
is a quantitative, not a qualitative question. Given \emph{any}
mechanism that selects the $\psi_2$ state from the $\Gamma_5$ manifold,
one expects a soft-mode in the spin-wave spectrum with a small
gap \cite{mcclarty-stasiak-gingras-2014-order}. Thus the observed excitation gap is a consequence of the $\psi_2$
selection and does not provide a signature for any given selection
mechanism. The identification of the mechanism thus lies in the
details, for example through the absolute magnitude of the gap or
dependence on other effects such as random disorder or magnetic field.
In the order by quantum disorder order scenario of
Ref. [\onlinecite{savary-balents-2012-order}], this final
identification rested on the effects of virtual crystal field
fluctuations being taken to be entirely negligible. As established in
Secs. \ref{sec:obvcef}, \ref{sec:biquadratic} and \ref{sec:results},
the assumptions unpinning this conclusion are unwarranted. Not only
are the effects of virtual crystal field fluctuations non-negligible,
they can be comparable to or \emph{even larger} than the corresponding
quantum fluctuations when multipolar interactions are taken into
account. We are thus left with the difficult task of
disentangling these two fluctuation effects, as both will be present
generically. 

We first should focus on the magnitude of the gap seen experimentally.
In Ref. [\onlinecite{ross-gaulin-2014-order}] and Ref.
[\onlinecite{petit-gingras-2014-order}] a gap of $\sim 40-45 \mueV$ is
observed. In the order by quantum disorder scenario, the excitation
gap is fully determined from the fitted exchanges of the effective
spin-1/2 model. However, this model is not analytically or numerically
tractable; given these theoretical limitations we must resort to
approximate methods to estimate the gap.  In
Ref. [\onlinecite{savary-balents-2012-order}], the large-$S$ result
from linear spin-wave theory (LSWT) is extrapolated to $S=1/2$ giving
a gap of $\Delta \sim 20 \mueV$, about $50\%$ of the observed gap
size.  A complementary perturbative approach undertaken in
Ref. [\onlinecite{maryasin-zhitomirsky-2014-order}] suggests that
effects beyond LSWT could further \emph{reduce} this value.  Using
real-space perturbation theory, Ref. [\onlinecite{maryasin-zhitomirsky-2014-order}] 
finds that the $U(1)$ breaking term obtained from LSWT is exactly canceled at higher-order in
perturbation theory when evaluated for $S=1/2$. A more careful
calculation \cite{maryasin-zhitomirsky-2014-order} gives a degeneracy
lifting $\delta E$ that is reduced by $\sim 40\%$ from the value
reported in Ref. [\onlinecite{savary-balents-2012-order}]. Since we
expect $\Delta^2 \sim C_2 \delta E$ (see Eq. (\ref{eq:energygap})),
this would reduce the gap size and move us significantly further away
from the experimental result.  If taken at face value, these estimates
suggest that the order by quantum-disorder contribution to the
condensation energy of the $\psi_2$ state could be as little as $25\%$
of the value inferred from the experimental gap. This raises some
doubt to whether order by quantum-disorder is the leading origin of
$\psi_2$ selection in \eto{}.

Another indirect probe into the selection mechanism is provided by the
effects of disorder. It has been argued
\cite{maryasin-zhitomirsky-2014-order,mcclarty-2015-order-dilution}
that diluting Er\tsup{3+} in \eto{} with non-magnetic Y\tsup{3+} ions
could provide another mechanism to break the degeneracy of the
$\Gamma_5$ manifold. Such dilution favors the $\psi_3$ state, not the
$\psi_2$ state found in the clean limit. This presents the intriguing
possibility that by a controlled introduction of vacancies at the
Er\tsup{3+} sites, one could tune from the $\psi_2$ state into the
$\psi_3$ state. The critical dilution would provide some indication of
condensation energy that stabilizes the $\psi_2$ state.  Estimating
this critical dilution, $\rho_c$, suffers from many of the theoretical
limitations such as estimating the gap in the pure effective spin-1/2
model. In Ref. [\onlinecite{maryasin-zhitomirsky-2014-order}], this is
estimated to be as low as $\rho_c \sim 7\%$, taking the degeneracy
lifting term $\delta E$ to be $\sim 40\%$ of the LSWT result, as
described above.  If instead of relying on the theoretical
stabilization energy, we extract $\delta E$ from the
\emph{experimental} gap using Eq. (\ref{eq:energygap}), one obtains a
$\delta E$ that is four times larger. An equivalent enhancement of
$\rho_c$ follows, raising the critical dilution to 25\% or so.  This
could provide further evidence towards a stronger selection effect
than predicted by the order by quantum disorder proposals.

The question then remains: what is responsible for the remaining
stabilization of the $\psi_2$ state?  The results presented here
suggest that this quantitative difference could be resolved through
the effects of virtual crystal field fluctuations. Further, we have
shown that the \emph{full} experimental gap can be accounted for with
reasonable values of the biquadratic coupling. Accounting for only the
fraction remaining after order by quantum disorder requires even less
biquadratic coupling.  We caution that a definitive experimental
signature of this order by virtual crystal field fluctuations
mechanism is difficult to formulate.  These questions are
highly quantitative and, given the small scale of these effects, are thus
muddled by the combination of theoretical and experimental
uncertainties. In principle, the presence of higher-rank multipolar
couplings could be probed by looking at the dispersion of the higher
crystal field levels in the ordered state. For example, a rough
estimate of the magnitude of the higher multipolar interactions could
be inferred by comparing with predictions assuming only the bilinear
interactions of Eq. (\ref{eq:bilinear-model}). However, given the
large number of independent parameters defining the
multipolar couplings, extracting a definite scale
for the higher-rank parts is likely to be an ill-posed challenge.
Another possible signature could be the finite $m_z$ moment. Indeed,
this is zero at leading order in the LSWT of
Ref. [\onlinecite{savary-balents-2012-order}] and in the real space
perturbation theory of
Ref. [\onlinecite{maryasin-zhitomirsky-2014-order}]. There are several
difficulties with this: it is generically expected to be finite for the
$\psi_2$ state, as it is allowed by symmetry \cite{javanparast-gingras-2015-order}, 
and thus should appear at
higher order in both such theories.  Further, the relationship between
the size of $m_z$ and the condensation energy is indirect, and even
when enhanced it may be too small to be observed experimentally.

\section{Conclusion}
\label{sec:conclusion}
The concept of order by disorder is a cornerstone of the theory of
highly frustrated magnetism \cite{gingras-mcclarty-2014-quantum,
villain-1980-order,henley-1989-ordering}, representing a middle ground between
ordering in an unfrustrated system and a fully magnetically disordered
state.  However, due to the need for a nearly accidental symmetry and
the difficulty in distinguishing it from more conventional ordering
scenario, clear material examples are scarce. Even in a compelling
material such as \eto{}, we have shown here that it is difficult to
distinguish between possible selection mechanisms. But not all interesting
physics is lost; we have
proposed a novel order by disorder mechanism that proceeds through the
virtual crystal field excitations of the Er\tsup{3+} ion. Despite the
naively large crystal field energy scale, this order by virtual
crystal field fluctuations can be quite effective in selecting the
ordered ground state. While the selection induced solely by bilinear
interactions is to some extent weaker than that seen experimentally (e.g.  the
gap, $\Delta$, is roughly five times too small), once multipolar
interactions are taken into account, we find this mechanism to be
competitive with order by quantum-disorder and is able to account for the
experimentally observed excitation gap, maintaining agreement
with the rest of the excitation spectrum. While determining the relative
size of the contributions of each of these mechanism is difficult, the
availability of high-quality single crystals of \eto{} provide some
hope that these effects could ultimately be disentangled.  We further
note that this selection mechanism is generically present in
rare-earth magnets, with its relevance primarily controlled by the gap
to the excited crystal field states.

To obtain a clearer signal for order by quantum- or thermal-disorder,
one should then look for materials with a larger crystal field
scale. A good candidate may be provided by \abo{Yb}{Ge} \cite{dun-hallas-2014-chemical} which
should be described using the same effective spin-1/2 model as \eto{} at low
energies \cite{
jaubert-gingras-2015-multiphase,robert-petit-2015-dynamics-competing}.
As the related material \abo{Yb}{Ti} appears to be proximate
to the $\Gamma_5$ phase \cite{jaubert-gingras-2015-multiphase,robert-petit-2015-dynamics-competing}, 
it is plausible that the ground state
of \abo{Yb}{Ge} is drawn from this manifold. All of the considerations
for \eto{} are then applicable, with the significant difference that the crystal
field energy scale in \abo{Yb}{Ge},
as in \yto{} \cite{bertin-2012-crystal},
 is likely an order of magnitude larger.  This renders the
order by virtual-crystal-field-fluctuations mechanism described in the
present work negligible, leaving for all practical 
purposes only quantum fluctuations to select
the ground state.

\acknowledgments{
  This work was supported by the NSERC of Canada, the Canada Research
  Chair program (M.G., Tier 1), the Canadian Foundation
  for Advanced Research and the Perimeter Institute (PI)
  for Theoretical Physics. Research at PI is supported by the
  Government of Canada through Industry Canada and by the
  Province of Ontario through the Ministry of Economic Development
  \& Innovation. M.G. acknowledges the hospitality and
  generous support of the Quantum Matter Institute at the University
  of British Columbia and TRIUMF where part of this
  work was completed.  
}

\appendix

\section{Local basis}
\label{basis}
We follow the conventions of \citet{savary-balents-2012-order} and work in the
basis local to each pyrochlore site. From the global basis, these
local axes are defined as
\begin{align}
  \vhat{z}_1 &= \frac{1}{\sqrt{3}} \left(+\vhat{x}+\vhat{y}+\vhat{z}\right), &
  \vhat{x}_1 &= \frac{1}{\sqrt{6}} \left(-2\vhat{x}+\vhat{y}+\vhat{z}\right),   \nonumber
  \\
  \vhat{z}_2 &= \frac{1}{\sqrt{3}} \left(+\vhat{x}-\vhat{y}-\vhat{z}\right), &
  \vhat{x}_2 &= \frac{1}{\sqrt{6}} \left(-2\vhat{x}-\vhat{y}-\vhat{z}\right),   \nonumber
  \\
  \vhat{z}_3 &= \frac{1}{\sqrt{3}} \left(-\vhat{x}+\vhat{y}-\vhat{z}\right), &
  \vhat{x}_3 &= \frac{1}{\sqrt{6}} \left(+2\vhat{x}+\vhat{y}-\vhat{z}\right),   \nonumber
  \\
  \vhat{z}_4 &= \frac{1}{\sqrt{3}} \left(-\vhat{x}-\vhat{y}+\vhat{z}\right), &
  \vhat{x}_4 &= \frac{1}{\sqrt{6}} \left(+2\vhat{x}-\vhat{y}+\vhat{z}\right), 
\end{align}
where $\vhat{y}_i = \vhat{z}_i \times \vhat{x}_i$.  The bond phase
factors $\gamma_{ij}$ and $\zeta_{ij} = -\cc{\gamma}_{ij}$ depend only
on the basis sites they connect and thus can be expressed as a matrix
\begin{equation}
  \gamma = \left(\begin{tabular}{cccc}
      $0$ & $+1$ & $\omega$ & $\omega^2$ \\
      $+1$ & $0$ & $\omega^2$ & $\omega$ \\
      $\omega$ & $\omega^2$ & $0$ & $+1$ \\
      $\omega^2$ & $\omega$ &$+1$ & $0$
      \end{tabular}\right), 
\end{equation}
where $\omega = e^{2\pi i/3}$.

\section{Model parameters}
\label{parameters}
\subsection{Crystal fields}
\label{crystal-field}
Several parametrizations of the crystal field potential for the
Er\tsup{3+} ion in \eto{} exist. We consider two of
these parameter sets: those of \citet{petit-gingras-2014-order} and
\citet{bertin-2012-crystal}. Both provide a good description of the
crystal field levels observed experimentally.  Considering the results
for these two parameters sets serve as a benchmark to the sensitivity
of our conclusions to the precise details of the crystal field
potential.  We write the crystal field potential $\cfV(\vec{\mJ})$ as
\begin{equation}
  \cfV(\vec{\mJ}) = \sum_{KQ} B_{KQ} \tilde{O}_{KQ}(\vec{\mJ}),
\end{equation}
where the $\tilde{O}_{KQ}(\vec{\mJ})$ are Stevens' operators \cite{stevens-1952-matrix}
defined using the conventions listed in \citet{jensen-1991-rare}.
Converted into the Stevens' convention, the non-zero parameters of
\citet{petit-gingras-2014-order} are
\begin{align}
  B_{20} &= +6.741 \cdot 10^{-2}\ \meV,&
  B_{40} &= +1.363 \cdot 10^{-3}\ \meV, \nonumber &\\
  B_{43} &= -8.998 \cdot 10^{-3}\ \meV, &
  B_{60} &= +9.565 \cdot 10^{-6}\ \meV, \nonumber & \\
  B_{63} &= +1.113 \cdot 10^{-4}\ \meV,& 
  B_{66} &= +1.661 \cdot 10^{-4}\ \meV.
\end{align}
The ground state $\lambda$-factors are given by $\lambda_{\pm} =
5.706$ and $\lambda_{z}=2.136$ and the first excited doublet lies at
$7.51\ \meV$.  The parameters of \citet{bertin-2012-crystal} are given
by
\begin{align}
  B_{20} &= +7.50 \cdot 10^{-2}\ \meV,&
  B_{40} &= +1.41 \cdot 10^{-3}\ \meV, \nonumber &\\
  B_{43} &= +1.25 \cdot 10^{-2}\ \meV, &
  B_{60} &= +1.09 \cdot 10^{-5}\ \meV, \nonumber & \\
  B_{63} &= -1.80 \cdot 10^{-4}\ \meV,& 
  B_{66} &= +1.50 \cdot 10^{-4}\ \meV.
\end{align}
The ground state $\lambda$-factors are given by $\lambda_{\pm} = 6.434
$ and $\lambda_{z}=1.758$ and the first excited doublet lies at $6.15\
\meV$.
\subsection{Exchanges}
\label{exchanges}
Estimates for the exchanges in the effective spin-1/2 model have been
extracted from fitting the excitation spectrum seen inelastic neutron
neutron scattering
experiments \cite{savary-balents-2012-order,petit-gingras-2014-order}. The fitting of
\citet{savary-balents-2012-order} was done in a magnetic field and yields the
parameters
\begin{align}
  \exJ_{zz}     &= -2.50 \pm 1.80 \cdot 10^{-2}\ \meV,   \nonumber \\
  \exJ_{\pm}    &= +6.50 \pm 0.75  \cdot 10^{-2}\ \meV,  \nonumber \\
  \exJ_{\pm\pm} &= +4.20 \pm 0.50  \cdot 10^{-2}\ \meV,   \nonumber  \\
  \exJ_{z\pm}   &= -0.88 \pm 1.50 \cdot 10^{-2}\ \meV.
\end{align}
The fitting of \citet{petit-gingras-2014-order} was done in zero-field and
yields (Model B)
\begin{align}
  \exJ_{zz}     &= -2.2 \pm 0.1 \cdot 10^{-2}\ \meV,   \nonumber \\
  \exJ_{\pm}    &= +6.0 \pm 0.1  \cdot 10^{-2}\ \meV,  \nonumber \\
  \exJ_{\pm\pm} &= +4.3 \pm 0.1  \cdot 10^{-2}\ \meV,   \nonumber  \\
  \exJ_{z\pm}   &= -1.5 \pm 0.1 \cdot 10^{-2}\ \meV.
\end{align}
\citet{petit-gingras-2014-order} also did a fit directly using an RPA
calculation within the full $J=15/2$ manifold.  Projected into the
effective spin-1/2 model this yields (Model A)
\begin{align}
  \exJ_{zz}     &= -0.84 \pm 0.1 \cdot 10^{-2}\ \meV,   \nonumber \\
  \exJ_{\pm}    &= +5.93 \pm 0.1  \cdot 10^{-2}\ \meV,  \nonumber \\
  \exJ_{\pm\pm} &= +4.61 \pm 0.1  \cdot 10^{-2}\ \meV,   \nonumber  \\
  \exJ_{z\pm}   &= +0.91 \pm 0.1 \cdot 10^{-2}\ \meV.
\end{align}
These three parameter sets are qualitatively compatible: each has
large positive in-plane exchanges $\exJ_{\pm}$ and $\exJ_{\pm\pm}$ with
smaller $\exJ_{zz}$ and $\exJ_{z\pm}$ couplings.

\section{Differences with \citet{petit-gingras-2014-order}}
\label{app:diff}
In \citet{petit-gingras-2014-order}, a theoretical gap size of $\sim
15 \mueV$ was reported, with an $\eta^{-1/2}$ dependence under crystal
field rescaling. This was computed through the same RPA approximation
discussed in Sec. \ref{sec:rpa}, but using a model that has only
bilinear interactions between the $\vec{\mJ}$ moments. These results
were computed using of a cutoff in the computation of the RPA
excitation spectrum; some of the highest lying crystal field states
were omitted from the calculation. This has significant effects on the
$\eta \rightarrow \infty$ limit, effectively removing the $U(1)$
degeneracy that should appear in the effective spin-1/2 model.  This
causes the erroneously large gap size and the incorrect $\eta^{-1/2}$
scaling reported in \citet{petit-gingras-2014-order}.
\section{Spherical vs Cartesian quadrupoles}
\label{cartesian}
Here we give the relationship between the Cartesian quadrupoles
of Eq. (\ref{quadrupoles}) and the spherical tensors defined in Eq. (\ref{multipole})
\begin{subequations}
\begin{align}
O_{2,0}(\vec{\mJ}) &= \frac{1}{12\sqrt{357}} Q^{zz},  \\
O_{2,\pm 1}(\vec{\mJ}) &= \pm \frac{1}{12\sqrt{357}} \left[ \sqrt{\frac{2}{3}} 
\left(Q^{xz} \pm i Q^{yz} \right) \right], \\
O_{2,\pm 2}(\vec{\mJ}) &= \frac{1}{12\sqrt{357}} \left[\sqrt{\frac{1}{6}}\left(
Q^{xx} - Q^{yy} \pm 2i Q^{xy}\right) \right].
\end{align}
\end{subequations}

\bibliography{draft}

\end{document}